\documentclass[11pt,american,twoside,a4paper]{article}

\usepackage[T1]{fontenc}
\usepackage[utf8]{inputenc}
\usepackage[american,provide=*]{babel}
\usepackage{a4wide}
\usepackage{fourier,bbm,bm}
\usepackage{BOONDOX-calo}
\usepackage{amsmath,amsfonts,amsthm,amssymb}
\usepackage{color,xcolor,graphicx}
\usepackage{latexsym,mathtools,leftindex}
\usepackage[shortlabels]{enumitem}
\usepackage{xargs,xspace,needspace,xparse,easyReview}
\usepackage[active]{srcltx}
\usepackage[colorlinks=true]{hyperref}
\usepackage{fancyhdr}
\usepackage[toc]{appendix}
\usepackage[normalem]{ulem}
\usepackage[style=cap-alpha]{biblatex}

\colorlet{darkblue}{blue!50!black}
\hypersetup{
	colorlinks,%
	citecolor=darkblue,%
	filecolor=red,%
	linkcolor=darkblue,%
	urlcolor=darkblue,%
	pdfnewwindow=true,%
	pdfstartview={FitH}
}

\numberwithin{equation}{section}
\theoremstyle{plain}
\newtheorem{theorem}{Theorem}[section]
  \AtBeginEnvironment{theorem}{\Needspace{5\baselineskip}}
  \newcommand\bet{\begin{theorem}} \newcommand\eet{\end{theorem}}
\newtheorem{proposition}[theorem]{Proposition}
  \AtBeginEnvironment{proposition}{\Needspace{5\baselineskip}}
  \newcommand\bep{\begin{proposition}} \newcommand\eep{\end{proposition}}
\newtheorem{lemma}[theorem]{Lemma}
  \AtBeginEnvironment{lemma}{\Needspace{5\baselineskip}}
  \newcommand\bel{\begin{lemma}} \newcommand\eel{\end{lemma}}
\newtheorem{corollary}[theorem]{Corollary}
  \AtBeginEnvironment{corollary}{\Needspace{5\baselineskip}}
  \newcommand\bec{\begin{corollary}} \newcommand\eec{\end{corollary}}
\theoremstyle{definition}
\newtheorem{definition}[theorem]{Definition}
  \AtBeginEnvironment{definition}{\Needspace{5\baselineskip}}
  \newcommand\bed{\begin{definition}} \newcommand\eed{\end{definition}}
\newtheorem{remark}{Remark}[section]
  \AtBeginEnvironment{remark}{\Needspace{5\baselineskip}}
  \newcommand\ber{\begin{remark}} \newcommand\eer{\end{remark}}
\newtheorem{remarks}[remark]{Remarks}
  \AtBeginEnvironment{remarks}{\Needspace{5\baselineskip}}
  \newcommand\bers{\begin{remarks}} \newcommand\eers{\end{remarks}}
\AtBeginEnvironment{quote}{\Needspace{5\baselineskip}}
\newcommand{\beq}{\begin{equation}} \newcommand{\eeq}{\end{equation}}

\newcounter{smallarabics}
\newenvironment{arabicenumerate}
{\begin{list}{{\normalfont\textbf{(\arabic{smallarabics})}}}
		{\usecounter{smallarabics}\setlength{\itemindent}{0cm}
			\setlength{\leftmargin}{5ex}\setlength{\labelwidth}{4ex}
			\setlength{\topsep}{0.75\parsep}\setlength{\partopsep}{0ex}
			\setlength{\itemsep}{0ex}}}
	{\end{list}}
\newcommand{\ben}{\begin{arabicenumerate}} \newcommand{\een}{\end{arabicenumerate}}

\newcounter{smallroman}
\newenvironment{romanenumerate}
{\begin{list}{{\normalfont\textbf{(\roman{smallroman})}}}
		{\usecounter{smallroman}\setlength{\itemindent}{0cm}
			\setlength{\leftmargin}{5ex}\setlength{\labelwidth}{4ex}
			\setlength{\topsep}{0.75\parsep}\setlength{\partopsep}{0ex}
			\setlength{\itemsep}{0ex}}}
	{\end{list}}
\newcommand{\benr}{\begin{romanenumerate}} \newcommand{\eenr}{\end{romanenumerate}}

\AtBeginEnvironment{theorem}{\Needspace{5\baselineskip}}
\AtBeginEnvironment{proposition}{\Needspace{3\baselineskip}}
\AtBeginEnvironment{definition}{\Needspace{5\baselineskip}}
\AtBeginEnvironment{corollary}{\Needspace{5\baselineskip}}
\AtBeginEnvironment{lemma}{\Needspace{5\baselineskip}}
\AtBeginEnvironment{quote}{\Needspace{5\baselineskip}}

\newcommandx{\ca}[2][1=]{\todo[inline,author={ca},
	linecolor=green,backgroundcolor=green!15,bordercolor=green,#1]{#2}}
\newcommandx{\canot}[2][1=]{\todo[author={ca},
	linecolor=green,backgroundcolor=green!15,bordercolor=green,#1]{#2}}

\NewDocumentCommand{\ASS}{mm}{\expandafter\newcommand\csname #1\endcsname{{\hyperref[#1]{\bf (#2)}}}}
\NewDocumentCommand{\preASS}{mm}{\expandafter\newcommand\csname pre#1\endcsname{{\hyperref[#1]{\bf (#2)}}}}

 \newcommand\cB{{\mathcal B}} \newcommand\cC{{\mathcal C}}
  \newcommand\cF{{\mathcal F}}
 \newcommand\cH{{\mathcal H}} 
  
 \newcommand\cN{{\mathcal N}} 
  
\newcommand\cS{{\mathcal S}} \newcommand\cT{{\mathcal T}}

\newcommand\NN{{\mathbbm N}}
\newcommand\ZZ{{\mathbbm Z}}

\newcommand\RR{{\mathbbm R}}
\newcommand\CC{{\mathbbm C}}

\newcommand\II{{\mathbbm I}}
\newcommand{\one}{\mathbbm{1}}

\newcommand\bY{\boldsymbol{Y}}

\newcommand\by{\boldsymbol{y}}

\newcommand\sS{{\mathsf S}}

\newcommand\fA{{\mathfrak A}}

\newcommand\fM{{\mathfrak M}}

 \newcommand\fh{{\mathfrak h}}

\newcommand\fs{{\mathfrak s}}

\newcommand\e{\mathrm{e}}

\renewcommand{\i}{\mathrm{i}}
\renewcommand\d{\mathrm{d}}

\newcommand{\Id}{\mathrm{Id}}

\newcommand{\argdot}{\,{\bm\cdot}\,}

\newcommand\myoperator[1]{\mathop{\mathrm{#1}}}

\newcommand\Dom{\myoperator{Dom}\nolimits}

\renewcommand\dim{\myoperator{dim}\nolimits}

\renewcommand\Im{\myoperator{Im}\nolimits}

\newcommand\tr{\myoperator{tr}\nolimits}

\newcommand{\Aut}{\myoperator{Aut}\nolimits}

\newcommand{\dist}{\myoperator{dist}\nolimits}
\newcommand\supp{\myoperator{supp}\nolimits}

\newcommand\wstar[1]{$\mathrm{w}^\ast\!\!\mathrm{-#1}$}

\renewcommand\proof{\noindent{\bf Proof.\ }}
\renewcommand\qed{\hfill$\Box$}
\newcommand\ie{\textsl{i.e.,\ }}
\renewcommand\bar{\overline}

\newcommand{\twol}[2]{\genfrac{}{}{0pt}{1}{#1}{#2}}

\newcommand\eq{\mathrm{eq}}

\newcommand\re{\mathrm{sa}}
\newcommand{\diam}{\mathop{\oslash}\nolimits}
\newcommand{\fAloc}{{{\mathfrak{A}}_{\mathrm{loc}}}}
\newcommand{\cBb}{\cB_\mathrm{b}}
\newcommand{\cBf}{\cB_\mathrm{f}}
\newcommand{\cBs}{\cB_\mathrm{s}}
\newcommand{\cBr}{\cB^r}
\newcommand{\cBsd}{\cB_\mathrm{sd}}
\newcommand{\cBdiam}[1]{\cB^{\diam^{#1}\!\!\!\!}}
\newcommand{\cBa}{\leftindex_\alpha\cB}
\newcommand{\cSI}{\cS_\mathrm{I}}
\newcommand{\cSeq}{\cS_\eq}
\newcommand{\cSwg}{\cS_\mathrm{wg}}

\newcommand{\eitHLP}{{\e^{\i tH_{\Lambda}(\Phi)}}}
\newcommand{\emitHLP}{{\e^{-\i tH_{\Lambda}(\Phi)}}}
\newcommand{\norm}[1]{{\|#1\|}}

\newcommand\Lim{\lim_\Lambda}
\newcommand\Llim{\lim_{\{\Lambda\}}}
\newcommand\fAr{\fA_\re}

\begin{document}
\renewcommand\today{}
\title{Dynamical Conservation Laws in Quantum  Spin Systems}

\author{V. Jak\v{s}i\'c$^{1}$,   C.-A. Pillet$^{2,3}$,  A. Szczepanek$^{4}$
\\ \\
$^{1}$Dipartimento di Matematica, Politecnico di Milano\\
piazza Leonardo da Vinci 32, 20133 Milano,  Italy
\\ \\
$^2$Université de Toulon, CNRS, CPT, UMR 7332, 83957, La Garde, France
\\ \\
$^3$Aix-Marseille Univ, CNRS, CPT, UMR 7332, Case 907, 13288, Marseille, France
\\ \\
$^4$Institute of Mathematics, Jagiellonian University \\
Łojasiewicza 6, 30-348 Kraków, Poland
}

\maketitle
\thispagestyle{empty}
\bigskip\noindent{\small{\bf Abstract.} We study dynamical conservation laws and
the propagation of thermodynamic structure in trans\-la\-tion-invariant lattice
quantum spin systems. Using Lieb--Robinson bounds
\href{https://doi.org/10.1007/BF01645779}{[CMP~28~(1972), 251--257]} in
arbitrary dimension, and Araki's bounds
\href{https://doi.org/10.1007/bf01645134}{[CMP~14~(1969), 120--157]} in
dimension one, we prove conservation of the specific entropy, specific energy,
and specific relative entropy under suitable assumptions on the interactions.  
The transported local Hamiltonians admit surface energies that vanish
per unit volume in the van~Hove limit, and local observables remain
a core for the transported generators at every finite time.
Weak Gibbsianity is preserved for sufficiently small times in arbitrary
dimension and for all times in dimension one. These results provide
the analytical foundation for the structural theory of approach to
equilibrium developed in a companion paper, and establish partial
results toward the conjectures formulated there.

}

\tableofcontents

\section{Introduction}
\label{sec:introduction-n}
This paper is a companion to~\cite{Jaksic2026a} and continues the program
initiated in~\cite{Jaksic2023a}. Throughout the paper, and without further
comment, we use the terminology, notation, and results of these works.

Two central ingredients of~\cite{Jaksic2023a} are:

\begin{description}
\item[\textbf{(a) Conservation laws.}] For $\Phi\in\cBr$, $\omega\in\cSI$,
and all $t\in\RR$,
\beq
s(\omega_t)=s(\omega),\qquad \omega_t(E(\Phi))=\omega(E(\Phi)),
\label{eq:claw1}
\eeq
where $\omega_t=\omega\circ\tau_\Phi^t$. The first relation
in~\eqref{eq:claw1} describes conservation of the specific entropy and goes
back to~\cite{Lanford1968}. The second one describes conservation of the
specific energy and was established in~\cite{Jaksic2023a}.

\item[\textbf{(b) Regular pairs.}] A pair
$(\omega,\Psi)\in\cSI\times\cBr$ is called \textsl{regular} if, for all
$\nu\in\cSI$, the specific relative entropy
\beq
s(\nu|\omega)\coloneq\Llim
\frac{S(\nu_\Lambda|\omega_\Lambda)}{|\Lambda|}
\label{eq:ad-Re}
\eeq
exists and satisfies
\beq
s(\nu|\omega)=-s(\nu)+\nu(E(\Psi))+P(\Psi),
\label{eq:ad-Re-1}
\eeq
see~\cite{Jaksic2023a}. This notion is related to some deep open problems in
quantum statistical mechanics. It is conjectured that, for any $\Psi\in\cBr$ and
$\omega\in\cSeq(\Psi)$, the pair $(\omega,\Psi)$ is regular.\footnote{The proof of
this conjecture published in~\cite{Moriya2020} is incorrect; see
footnote~11 in~\cite{Jaksic2023a}.}  A stronger conjecture is that, for any
$\Psi\in\cBr$, one has
\beq
\cSwg(\Psi)=\cSeq(\Psi),
\label{eq:weq}
\eeq
where $\cSwg(\Psi)$ is the set of weak Gibbs states introduced
in~\cite[Section~2.2]{Jaksic2023a}.\footnote{For
$\omega\in\cSwg(\Psi)$, the existence of the limit~\eqref{eq:ad-Re} and
the validity of~\eqref{eq:ad-Re-1} are automatic.} Although the inclusion
$\cSwg(\Psi)\subset\cSeq(\Psi)$ is immediate, the equality~\eqref{eq:weq}
is only known to hold for one-dimensional finite-range interactions and in
the high-temperature regime $\|\Psi\|_r<r$ in arbitrary dimension; see
\cite[Theorem~2.11]{Jaksic2023a}.
\end{description}

Starting with~(a) and~(b), \cite{Jaksic2023a} develops the structural theory
of Dynamical Steady States (DSS) in lattice quantum spin and fermion
systems, with a focus on the phenomenon of \textsl{approach to
equilibrium}.\footnote{Often called the \textsl{zeroth} or
\textsl{minus first law of thermodynamics}; see~\cite{Brown2001}.}
The companion paper~\cite{Jaksic2026a} further develops this theory by
introducing physically relevant constants of motion and establishing a
\textsl{dynamical Gibbs variational principle} for DSS. When the quantum Stein lemma applies, this principle identifies the
physical equilibrium states describing a DSS as those least
distinguishable from the initial state.
For DSS admitting a physical equilibrium description, the theory gives
a precise mathematical form to the idea that the dynamics selects states
of maximal specific entropy subject to the constraints imposed by the
constants of motion. 

A central large-time question in the framework developed there is
restoration of locality: whether the DSS admit a regular local
thermodynamic description.
This structural picture rests on a number of properties of quantum spin
systems which are either kinematical or concern  finite-time dynamics,
and hence are logically prior to the large-time problem of approach to
equilibrium. In~\cite{Jaksic2026a}, several of these properties are
formulated as conjectures. The present paper establishes finite-time conservation and propagation
results that support this structural picture and provide partial progress
toward these conjectures. Proving the conjectures in full generality
remains a substantial open problem in quantum statistical mechanics.

Our focus will be on:

\begin{description}
\item[\textbf{(1)}] The conservation laws~\eqref{eq:claw1} for
interactions in diameter-based Banach spaces. These spaces are not
comparable with the spaces $\cB^r$ used in the earlier proofs
of~\cite{Lanford1968,Jaksic2023a}. The proof of conservation of the
specific entropy follows the approach of~\cite{Wreszinski2019,Wreszinski2022},
using Lieb--Robinson bounds.

\item[\textbf{(2)}] Properties of the families
$(H_\Lambda(\Psi|\Phi,t))_{\Lambda\in\cF}$ of evolved local Hamiltonians,
$$
H_\Lambda(\Psi|\Phi,t)
=
\e^{-\i tH_\Lambda(\Phi)}
H_\Lambda(\Psi)
\e^{\i tH_\Lambda(\Phi)},
$$
and of the dynamics they generate in the thermodynamic limit. This involves,
in particular, the control of the associated surface energies
$W_\Lambda(\Psi|\Phi,t)$, the propagation from $t=0$ to finite times of the
property
$$
\Lim\frac{\|W_\Lambda(\Psi|\Phi,t)\|}{|\Lambda|}=0,
$$
and preservation of the Gibbs condition for $\omega_t$ with respect
to $H(\Psi|\Phi,t)$. We establish preservation of weak Gibbsianity
under additional assumptions. We also prove that local observables
remain a core for the generator of the transported dynamics at
every finite time under suitable decay assumptions.
In the high-temperature regime, the $\cB^r$ condition does not by itself
provide the spatial decay required for Lieb--Robinson bounds. The
preservation of weak Gibbsianity is therefore proved by refining Ruelle's
analytic estimates~\cite{Ruelle1969}, while the result for all times in dimension one is obtained by
adapting Araki's bounds~\cite{Araki1969}.

\item[\textbf{(3)}] Verification of
\cite[Condition~\textbf{(SD+)}]{Jaksic2026a} under suitable decay
assumptions. The characterization proved there then implies that
$E(\Psi)$ is a constant of motion for $\tau_\Phi$ if and only if
the dynamics $\tau_\Phi$ and $\tau_\Psi$ commute.

\item[\textbf{(4)}] Conservation of the specific relative entropy
$s(\nu_t|\omega_t)$. For finite-range $\Phi$ and a regular pair
$(\omega,\Psi)$, we prove this conservation law in two regimes: for
sufficiently small times at high temperature in arbitrary dimension,
and for all times in dimension one when $\Psi$ has finite range.
We extend the notion of regularity from interactions to families of
local Hamiltonians and show that, in these regimes, the evolved state
$\omega_t$ is regular with respect to $H(\Psi|\Phi,t)$. Under weaker assumptions on the interactions, we use the Hiai--Petz
bound to show that, for a regular pair $(\omega,\Psi)$, the specific
relative entropy $s(\nu_t|\omega_t)$, whenever it exists, does not
exceed $s(\nu|\omega)$.

\end{description}

Although we restrict ourselves to lattice quantum spin systems, the arguments
extend, with minor adaptations, to lattice fermions. Beyond their role in the
companion paper~\cite{Jaksic2026a}, the results above are of independent
interest as conservation and propagation results for infinite quantum
systems.

The paper is organized as follows. In Section~\ref{sec:Setup} we briefly
introduce the general setup of lattice quantum spin systems and some
additional notation.\footnote{Familiarity with the notation and terminology
of~\cite{Jaksic2023a} is assumed.} In Section~\ref{sec:main-Res} we state
our main results, whose proofs are given in Section~\ref{sec:Proofs}.

\paragraph*{Acknowledgments.}
This work was partly funded by the CY Initiative grant
\emph{Investissements d'Avenir}, grant number ANR-16-IDEX-0008.
VJ acknowledges the support of the MUR grant
``Dipartimento di Eccellenza 2023--2027'' of the Dipartimento di Matematica,
Politecnico di Milano.
We also acknowledge the support of the ANR project DYNACQUS,
grant number ANR-24-CE40-5714. 
AS is grateful to Clément Tauber for helpful discussions during the early stages of the project.

\section{Setup}
\label{sec:Setup}
We consider translation-invariant quantum spin systems on $\ZZ^d$,
with distance $|x-y|$, where $|x|\coloneq\sum_{j=1}^d|x_j|$.
Let $\cF$ denote the collection of finite subsets of $\ZZ^d$.
For $X\in\cF$, $|X|$ denotes its cardinality and, when $X\ne\emptyset$,
$\diam(X)\coloneq\max_{x,y\in X}|x-y|$ its diameter.
For any $X\subset\ZZ^d$,  $X^c\coloneq\ZZ^d\setminus X$.

A net $(\Lambda_\alpha)_{\alpha\in I}$ in $\cF$ is called a
\emph{van~Hove net} if it is exhaustive and, for every $r>0$,
$$
\lim_{\alpha\in I}\frac1{|\Lambda_\alpha|}
\left|
\left\{
x\in\Lambda_\alpha\mid
\dist(x,\Lambda_\alpha^c)\leq r
\right\}
\right|
=0.
$$
A function $f$ on $\cF$ with values in a metric space is said to have
the van~Hove limit $L$ if
$$
\lim_{\alpha\in I}f(\Lambda_\alpha)=L
$$
for every van~Hove net $(\Lambda_\alpha)_{\alpha\in I}$. We write $\Lim f(\Lambda)=L$. 
We shall also use limits along the directed family of cubes centered at the
origin, ordered by inclusion, and write $\Llim f(\Lambda)$
for such a limit. Since centered cubes form a van~Hove net, the existence of
$\Lim f(\Lambda)$ implies that $\Llim f(\Lambda)$ exists and has the same
value. The converse need not hold.

Although the van~Hove limit is conceptually more intrinsic,\footnote{We
refer to the Appendix of~\cite{Araki2003} for a detailed exposition of the
van~Hove limit.} it is often technically more convenient to work along
cubes. As in~\cite{Jaksic2023a}, unless explicitly stated otherwise,
limits in which a finite region exhausts $\ZZ^d$ are taken along centered
cubes.  Van~Hove limits enter Theorem~\ref{thm:W_disapp} and its proof.
They are required by~\cite[Condition~\textbf{(SD)}]{Jaksic2026a}
and by the Araki--Moriya formalism used in the proof of
\cite[Theorem~4.8]{Jaksic2026a} under \textbf{(SD+)}.

We shall also take limits over the finite supersets $\widetilde\Lambda$
of a fixed $\Lambda\in\cF$, directed by inclusion. These limits enter
the definition of surface energies in the Araki--Moriya
formalism~\cite{Araki2003}. They are distinct from the thermodynamic
limits discussed above: no van~Hove condition is imposed on
$\widetilde\Lambda$.

For any $X\in\cF$, $\cH_X\coloneq\bigotimes_{x\in X}\cH_x$, where $\cH_x=\cH$
is the finite-dimensional Hilbert space of a single spin. The spin algebra over
$X$ is $\fA_X\coloneq\cB(\cH_X)$, where $\cB(\cH)$ denotes the $C^\ast$-algebra
of linear operators on $\cH$. We denote by $\one_X$ the unit of $\fA_X$. If
$X\subset Y\in\cF$, then $A\in\fA_X$ is identified with
$A\otimes\one_{Y\setminus X}\in\fA_Y$, so that $\fA_X\subset\fA_Y$. The spin
algebra $\fA$ is the inductive limit of the family $(\fA_X)_{X\in\cF}$ of
$C^\ast$-algebras. Elements of $\fAloc\coloneq\bigcup_{X\in\cF}\fA_X$ are said
to be local, and form a dense $\ast$-subalgebra of $\fA$. We denote by $\fAr$
the set of self-adjoint elements of $\fA$, by $\cS$ the set of states on $\fA$
equipped with its \wstar{topology}, and by $\Aut(\fA)$ the group of its
$\ast$-automorphisms. Let $(\fh_\omega,\pi_\omega,\Omega_\omega)$ be the
GNS-representation of $\fA$ induced by the state $\omega\in\cS$. The folium
of this state is the subset $\cN_\omega\subset\cS$ of states of the form
$\nu(A)=\tr(\rho\pi_\omega(A))$, where $\rho$ is a density matrix on
$\fh_\omega$, \ie a positive trace class operator on $\fh_\omega$ of unit trace. 
 A $C^\ast$-dynamics on $\fA$ is a strongly continuous one-parameter subgroup
$\tau=\left(\tau^t\right)_{t\in\RR}$ of $\Aut(\fA)$. We denote by $\ZZ^d\ni
x\mapsto\varphi^x\in\Aut(\fA)$ the natural group action of $\ZZ^d$ on $\fA$, and
by $\cSI$ the set of translation-invariant states on $\fA$.

\subsection{Interactions}
\label{sec:Inters}

An interaction is a family $\Phi\coloneq\left(\Phi(X)\right)_{X\in\cF}$, where
each $\Phi(X)\in\fA_X$ is self-adjoint, with the convention
$\Phi(\emptyset)=0$. We only consider translation-invariant interactions
satisfying $\varphi^x(\Phi(X))=\Phi(X+x)$ for all $x\in\ZZ^d$ and $X\in\cF$.
An interaction $\Phi$ is said to have finite range $R$ whenever
$$
R=\sup\{\diam(X)\mid X\in\cF,\Phi(X)\neq0\}<\infty.
$$
We set $R=0$ when $\Phi=0$, and  denote by $\cBf$ the real vector space of finite-range interactions.

We shall make use of several Banach spaces of interactions, each obtained by
completing $\cBf$ in a suitable norm. The big and small spaces,
$\cBb$ and $\cBs$, are associated with the norms
$$
\norm{\Phi}_\mathrm{b}\coloneq\sum_{X\ni0}\dfrac{\norm{\Phi(X)}}{|X|},
\qquad
\norm{\Phi}_\mathrm{s}\coloneq\sum_{X\ni0}\norm{\Phi(X)},
$$
respectively. The spaces $\cB^r$ and $\cBa$ correspond to norms with exponential
and polynomial weights in $|X|$, respectively:
$$
\norm{\Phi}_r\coloneq\sum_{X\ni0}\e^{r(|X|-1)}\|\Phi(X)\|,
\qquad
\leftindex_\alpha\|\Phi\|\coloneq\sum_{X\ni0}|X|^{\alpha}\norm{\Phi(X)},
$$
with $r>0$ and $\alpha\geq-1$. For $\alpha\geq0$, one has
$\cBf\subset\cBr\subset\cBa\subset\cBs\subset\cBb$, and for
$-1\leq\alpha\leq0$ one has
$\cBf\subset\cBr\subset\cBs\subset\cBa\subset\cBb$.
 Notice that, for
convenience, the parameter $\alpha$ in $\cBa$ has been shifted by $1$ when
compared to~\cite[Definition~2.37]{Van_Enter1993}.

The spaces $\cBdiam{\gamma}$ associated with the norms
$$
\|\Phi\|_{\diam^\gamma}\coloneq
\sum_{X\ni0}|X|(1+\diam(X))^\gamma\|\Phi(X)\|,
\qquad \gamma>0,
$$
will play an important role in this paper. Note that
$\cBf\subset\cBdiam{\gamma}\subset\cBs\subset\cBb$ and
$\cBdiam{\gamma}\subset\cBa$ for all $\gamma>0$ and $-1\leq \alpha\leq1$.
However, the spaces $\cBdiam{\gamma}$ and $\cBr$ are not comparable.

The \textsl{local Hamiltonians} associated with $\Phi$ are defined by
$$
H_{\Lambda}(\Phi)\coloneq\sum_{X\subset\Lambda}\Phi(X)\in\fA_\Lambda.
$$
The corresponding \textsl{local dynamics} are the one-parameter groups of
inner $\ast$-automorphisms of $\fA$ given by the usual Heisenberg evolution of
quantum mechanics
$$
\tau^t_{\Phi,\Lambda}(A)\coloneq\eitHLP A\emitHLP.
$$
For $\Phi\in\cBs$, let $\delta_\Phi$ be the $\ast$-derivation on $\fAloc$
defined by
$$
\delta_\Phi(A)=\sum_{X\cap\supp A\neq\emptyset}\i[\Phi(X),A].
$$
The derivation $\delta_\Phi$ is norm-closable, and we denote its closure by the
same symbol. Its closure generates a $C^\ast$-dynamics on $\fA$ if and only if
$(\Id\pm\delta_\Phi)\fAloc$ is dense in $\fA$; see
\cite[Theorem~3.2.50 and Proposition~3.2.52]{Bratteli1987} and
\cite[Proposition~6.2.3]{Bratteli1981}. We denote by $\cBsd$ the set of
interactions $\Phi\in\cBs$ for which this condition holds. For
$\Phi\in\cBsd$, the resulting dynamics is denoted by
$\tau_\Phi=(\tau_\Phi^t)_{t\in\RR}$. In this case,
$$
\tau^t_\Phi(A)=\Llim\tau^t_{\Phi,\Lambda}(A)
$$
for all $A\in\fA$, uniformly for $t$ in compact subsets of $\RR$.

It is a classical result~\cite[Theorem~6.2.4]{Bratteli1981} that
$\cB^r\subset\cBsd$. We shall also rely on the following consequence of the
Lieb--Robinson theory developed in~\cite{Nachtergaele2006,Nachtergaele2019};
see Section~\ref{sec:proof_bound_Lieb_Rob} for the proof of the estimate.

\bet\label{thm:bound_Lieb_Rob}
If $\gamma>d$, then $\cBdiam{\gamma}\subset\cBsd$. Moreover, for any
$\Phi\in\cBdiam{\gamma}$, $0<\delta<\gamma-d$ and $t\in\RR$, there exists
a constant $c_{\Phi,\delta,t}$ such that
$$
\left\|\tau_\Phi^t(A)-\tau^t_{\Phi,\Lambda}(A)\right\|
\leq
c_{\Phi,\delta,t}
\frac{|\Lambda'|}
{\left(1+\dist(\Lambda',\Lambda^c)\right)^{\delta}}
\|A\|,
$$
provided $A\in\fA_{\Lambda'}$ with $\Lambda'\subset\Lambda\in\cF$.
\eet

The \textsl{specific energy} of an interaction $\Phi\in\cBb$ is the observable
$$
E(\Phi)\coloneq\sum_{X\ni0}\frac{\Phi(X)}{|X|}\in\fA.
$$
It satisfies
\beq
\nu(E(\Phi))=\Llim\frac1{|\Lambda|}\nu(H_\Lambda(\Phi))
\label{eq:spec_en_ham},
\eeq
for any $\nu\in\cSI$; see~\cite[Proposition~6.2.39]{Bratteli1981}.

Given $\omega\in\cSI$ and $\Lambda\in\cF$, the restriction of $\omega$ to
$\fA_\Lambda$ is described by the density matrix
$\omega_\Lambda\in\fA_\Lambda$ such that\footnote{We will always identify a
density matrix in $\fA_\Lambda$ with the state it induces on $\fA_\Lambda$ in
this way.}
$$
\omega(A)=\tr_{\cH_\Lambda}(\omega_\Lambda A)
$$
for all $A\in\fA_\Lambda$. The \textsl{specific entropy} of $\omega$ is given by
$$
s(\omega)\coloneq\Llim\frac1{|\Lambda|}S(\omega_\Lambda),
$$
where
$S(\omega_\Lambda)=-\tr_{\cH_\Lambda}(\omega_\Lambda\log\omega_\Lambda)$
is the von~Neumann entropy of $\omega_\Lambda$. This limit always exists and
defines an affine, \wstar{upper} semicontinuous map
$s:\cSI\to[0,\log\dim\cH]$;
see~\cite[Proposition~6.2.38]{Bratteli1981}.

The \textsl{pressure} (or \textsl{free energy}) of $\Phi$,
$$
P(\Phi)
\coloneq
\Llim\frac1{|\Lambda|}
\log\,\tr_{\cH_\Lambda}\left(\e^{-H_\Lambda(\Phi)}\right)
$$
defines a convex $1$-Lipschitz function on $\cBb$ with respect to
$\|\argdot\|_\mathrm{b}$, satisfying the Gibbs variational principle
$$
P(\Phi)=\sup_{\omega\in\cSI}\left(s(\omega)-\omega(E(\Phi))\right),
$$
see~\cite[Theorem~6.2.40]{Bratteli1981}. The set of
\emph{equilibrium states} of $\Phi$ is
$$
\cSeq(\Phi)\coloneq\left\{\omega\in\cSI\mid
s(\omega)=\omega(E(\Phi))+P(\Phi)\right\}.
$$

Given a $C^\ast$-dynamics $\tau$ on $\fA$ and $\beta\in\RR\setminus\{0\}$,
$\omega\in\cS$ is a $(\tau,\beta)$\textsl{-KMS state} whenever, for all
$A,B\in\fA$, there exists an analytic function $F_{A,B}$ on the complex strip
$\sS_\beta\coloneq\{z\mid0\wedge\beta<\Im z<0\vee\beta\}$, which has a bounded
continuous extension to the closure of $\sS_\beta$, and such that for any
$t\in\RR$,
$$
F_{A,B}(t)=\omega(A\tau^t(B)),\qquad
F_{A,B}(t+\i\beta)=\omega(\tau^t(B)A).
$$
For $\beta=0$, a $(\tau,0)$-KMS state is a $\tau$-invariant tracial state.

For $\Phi\in\cBsd$ and $\beta\in\RR$, the set of all translation-invariant
$(\tau_\Phi,\beta)$-KMS states coincides with $\cSeq(\beta\Phi)$; for
$\beta\neq0$ this follows from~\cite[Theorems~A and~B]{Araki2003}, while
for $\beta=0$ both sets consist of the unique tracial state on $\fA$.

Let $\alpha$ be the $C^\ast$-dynamics generated by the $\ast$-derivation
$\delta$ of $\fA$, \ie $\alpha^t=\e^{t\delta}$. To any self-adjoint
$V\in\fA$ we associate the perturbed $C^\ast$-dynamics $\alpha_V$ generated
by $\delta_V=\delta+\i[V,\argdot]$. According to Araki's perturbation theory,
see~\cite[Section~5.4.1]{Bratteli1981}, to any $(\alpha,\beta)$-KMS state
$\omega\in\cS$ one canonically associates an $(\alpha_V,\beta)$-KMS state
$\omega_V\in\cN_\omega$. Moreover, the map $\fAr\ni V\mapsto\omega_V\in\cN_\omega$ is norm   continuous. 
We call $\omega_V$ the \textsl{perturbation of $\omega$ by $V$}.

A state $\omega\in\cS$ is \textsl{modular} whenever its GNS-vector
$\Omega_\omega$ is cyclic for the commutant $\pi_\omega(\fA)'$. Such a state
has a unique normal extension
$\widehat\omega(\argdot)=\langle\Omega_\omega,\argdot\,\Omega_\omega\rangle$
on the enveloping von~Neumann algebra $\fM_\omega=\pi_\omega(\fA)''$.
$\widehat\omega$ is $(\sigma_\omega,-1)$-KMS for a unique dynamics
$\sigma_\omega$, the modular group of $\omega$;
see~\cite[Theorem~5.3.10]{Bratteli1981}. In particular, for $\beta\neq0$,
$\widehat\omega$ is $(\tau,\beta)$-KMS for the dynamics
$\tau^t=\sigma_\omega^{-t/\beta}$ on $\fM_\omega$. Conversely, any state $\omega\in\cS$
which is $\beta$-KMS for some $C^\ast$-dynamics $\tau$ is modular, and its
modular group satisfies
$\sigma_\omega^t\circ\pi_\omega=\pi_\omega\circ\tau^{-\beta t}$.
Whenever $\omega$ is modular and $V\in\fAr$, we define $\omega_V$
by perturbing $\widehat\omega$, regarded as a
$(\sigma_\omega^{-\argdot},1)$-KMS state, by $\pi_\omega(V)$,
and pulling the resulting state back to $\fA$ through $\pi_\omega$.
This agrees with the preceding convention when $\beta=1$.

For $\Phi\in\cBs$, the surface energy associated with $\Lambda\in\cF$ is
$$
W_\Lambda(\Phi)
=
\sum_{\twol{X\cap\Lambda\neq\emptyset}
{X\cap\Lambda^c\neq\emptyset}}
\Phi(X).
$$
By absolute convergence, it can equivalently be written as the norm limit
\beq
W_\Lambda(\Phi)
=
\lim_{\widetilde\Lambda}W_{\Lambda,\widetilde\Lambda}(\Phi),
\qquad
W_{\Lambda,\widetilde\Lambda}(\Phi)
=
H_{\widetilde\Lambda}(\Phi)
-
H_\Lambda(\Phi)
-
H_{\widetilde\Lambda\setminus\Lambda}(\Phi),
\label{eq:surfacePsi}
\eeq
where the limit is over the finite supersets $\widetilde\Lambda$ of
$\Lambda$, directed by inclusion. We recall that
$$
\Lim\frac1{|\Lambda|}\|W_\Lambda(\Phi)\|=0,
$$
see Remark~(1) after~\cite[Proposition~6.2.41]{Bratteli1981}.

A state $\omega\in\cSI$ satisfies the \textsl{Gibbs condition} with respect to
$\Phi\in\cBs$ whenever it is modular and, for every $\Lambda\in\cF$, the
perturbed state $\omega_{-W_\Lambda(\Phi)}$ satisfies
$$
\omega_{-W_\Lambda(\Phi)}
=
\omega_{\Phi,\Lambda}\otimes\widetilde\omega_{\Lambda^c},
$$
where
$\omega_{\Phi,\Lambda}
=\e^{-H_\Lambda(\Phi)}/\tr(\e^{-H_\Lambda(\Phi)})$
is the local Gibbs state on $\fA_\Lambda$, and
$\widetilde\omega_{\Lambda^c}$ is some  state on
$\fA_{\Lambda^c}$.

A state $\omega\in\cSI$ is \textsl{weak Gibbs} with respect to $\Phi\in\cBb$,
\cite[Definition~2.9]{Jaksic2023a}, if there exist constants
$C_\Lambda>1$ such that $\Llim\frac1{|\Lambda|}\log C_\Lambda=0$ and
$$
{C_\Lambda}^{-1}\omega_{\Phi,\Lambda}
\leq
\omega_\Lambda
\leq
C_\Lambda\omega_{\Phi,\Lambda}
$$
for every cube $\Lambda$. We denote by $\cSwg(\Phi)$ the set of weak Gibbs states with respect to
$\Phi$. The proof of~\cite[Proposition~2.10]{Jaksic2023a} shows that
$\cSwg(\Phi)\subset\cSeq(\Phi)$.

\subsection{Local Hamiltonians}
\label{sec:LocHam}
We shall also work directly with families of local Hamiltonians
$H\coloneq(H_\Lambda)_{\Lambda\in\cF}$.
 For each $\Lambda\in\cF$, $H_\Lambda$ is a self-adjoint operator
on $\cH_\Lambda$. We also assume translation invariance, \ie
$\varphi^x(H_\Lambda)=H_{\Lambda+x}$ for each $x\in\ZZ^d$. The basic notions
introduced above extend naturally to this framework.
\benr
\item $H$ \emph{generates a dynamics} $\tau_H$ if the limit
$$
\tau_H^t(A)\coloneq
\Llim\tau_{H,\Lambda}^t(A),\qquad
\tau_{H,\Lambda}^t(A)\coloneq\e^{\i tH_\Lambda}A\e^{-\i tH_\Lambda},
$$
exists for all $A\in\fAloc$, uniformly for $t$ in compact intervals. In this
case, the family
$\tau_H\coloneq\left(\tau_H^t\right)_{t\in\RR}$ extends uniquely to a
$C^\ast$-dynamics on $\fA$.

\item The \emph{pressure} of $H$ is defined by
$$
P_H\coloneq\Llim\frac1{|\Lambda|}\log\,\tr_{\cH_\Lambda}\e^{-H_\Lambda},
$$
provided that this limit exists.

\item An observable $E_H\in\fA$ is a \emph{specific energy} of $H$ if
$$
\omega(E_H)=\Llim\frac1{|\Lambda|}\omega(H_\Lambda)
$$
for every $\omega\in\cSI$.

\item Whenever the pressure and specific energy  of $H$ exist, its
\emph{equilibrium states} are the elements of the set
$$
\cSeq(H)\coloneq\{\omega\in\cSI\mid s(\omega)=\omega(E_H)+P_H\}.
$$

\item Whenever the pressure and specific energy of $H$ exist, a pair
$(\omega,H)$ with $\omega\in\cSI$ is called \emph{regular} if, for every
$\nu\in\cSI$, the specific relative entropy $s(\nu|\omega)$ exists and
$$
s(\nu|\omega)=-s(\nu)+\nu(E_H)+P_H.
$$

\item The \emph{surface energy} of $H$ is the family
$W=(W_\Lambda)_{\Lambda\in\cF}$ defined by
\beq
W_\Lambda
\coloneq
\lim_{\widetilde\Lambda}
W_{\Lambda,\widetilde\Lambda},
\qquad
W_{\Lambda,\widetilde\Lambda}
\coloneq
H_{\widetilde\Lambda}
-
H_\Lambda
-
H_{\widetilde\Lambda\setminus\Lambda},
\label{eq:surface_en_def}
\eeq
provided these norm limits exist; the limits are taken over the finite supersets
$\widetilde\Lambda$ of $\Lambda$, directed by inclusion.

\item Assuming the existence of the surface energy $W$, a state
$\omega\in\cS$ satisfies the
\textsl{Gibbs condition with respect to $H$} if it is modular and, for
every $\Lambda\in\cF$, the perturbed state $\omega_{-W_\Lambda}$
satisfies
$$
\omega_{-W_\Lambda}
=
\omega_{H,\Lambda}\otimes\widetilde\omega_{\Lambda^c},
$$
where
$$
\omega_{H,\Lambda}
\coloneq
\frac{\e^{-H_\Lambda}}{\tr(\e^{-H_\Lambda})}
$$
is the local Gibbs state on $\fA_\Lambda$ associated with $H$, and
$\widetilde\omega_{\Lambda^c}$ is some state  on
$\fA_{\Lambda^c}$.

\item A state $\omega\in\cSI$ is \emph{weak Gibbs} with respect to
$H$ if there exist constants $C_\Lambda>1$ such that
$$
{C_\Lambda}^{-1} \omega_{H,\Lambda}\leq\omega_\Lambda\leq C_\Lambda\omega_{H,\Lambda}
$$
for every cube $\Lambda$ and
$$
\Llim\frac1{|\Lambda|}\log C_\Lambda=0.
$$
We denote by $\cSwg(H)$ the set of weak Gibbs states with respect to $H$.
Whenever the pressure and specific energy of $H$ exist, a straightforward
adaptation of the proof of~\cite[Proposition~2.10]{Jaksic2023a} shows that
$\cSwg(H)\subset\cSeq(H)$.
\eenr

\medskip
Of particular interest for our purposes is the family
$H(\Psi|\Phi,t)=\left(H_\Lambda(\Psi|\Phi,t)\right)_{\Lambda\in\cF}$
defined by
\beq
H_\Lambda(\Psi|\Phi,t)\coloneq\tau^{-t}_{\Phi,\Lambda}(H_\Lambda(\Psi))
=\e^{-\i tH_\Lambda(\Phi)}H_\Lambda(\Psi)\e^{\i tH_\Lambda(\Phi)},
\label{eq:HamHt}
\eeq
where $t\in\RR$ and $\Phi,\Psi$ are two interactions. The following proposition collects some elementary properties of this
family. Its proof is given in Section~\ref{sec:proofPropHt_DS}.

\bep\label{prop:Ht_SDS}
For any $\Psi\in\cB_\mathrm{b}$ and $t\in\RR$, the following hold:
\benr
\item If $\Phi\in\cB_\mathrm{b}$, then the pressure of $H(\Psi|\Phi,t)$
exists and coincides with the pressure of $\Psi$.
\item If either $\Phi\in\cBdiam{\gamma}$ for some $\gamma>2d$ and
$\Psi\in\cBs$, or $\Phi\in\cBdiam{\gamma}$ for some $\gamma>d$ and
$\Psi\in\leftindex_1\cB$, then the specific energy  of
$H(\Psi|\Phi,t)$ exists and is given by
$$
E_{H(\Psi|\Phi,t)}=\tau_\Phi^{-t}(E(\Psi)).
$$
\item If the local Hamiltonian families $H(\Phi)$ and $H(\Psi)$ generate
dynamics $\tau_\Phi$ and $\tau_\Psi$, then $H(\Psi|\Phi,t)$ generates the
dynamics
$$
\alpha_{H(\Psi|\Phi,t)}^s=\tau^{-t}_\Phi\circ\tau^s_\Psi\circ\tau^t_\Phi,
$$
for $s,t\in\RR$.
\eenr
\eep

\section{Main results}
\label{sec:main-Res}

\subsection{Conservation of specific entropy and energy}

Our first result states that the specific entropy and energy remain constant under
the dynamics generated by a large class of interactions.

\bet\label{thm:entropy_energy_cons}
Let $\Phi\in\cBdiam{\gamma}$, $\omega\in\cSI$, and $t\in\RR$.
\benr
\item If $\gamma>2d$, then
$$
s(\omega\circ\tau_\Phi^t)=s(\omega).
$$
\item If $\gamma>d$, then
$$
(\omega\circ\tau_\Phi^t)(E(\Phi))=\omega(E(\Phi)).
$$
\eenr
\eet

For $\Phi\in\cB^r$, conservation of the specific entropy and specific
energy was established in~\cite[Theorem~5]{Lanford1968}
and~\cite[Theorem~2.8]{Jaksic2023a}, respectively. The argument for
specific-energy conservation applies more generally to all
$\Phi\in\cBsd$. Here we work with the spaces $\cBdiam{\gamma}$,
which are not comparable with $\cB^r$, and prove both parts using
Lieb--Robinson bounds; see Section~\ref{sec:proof_entropy_energy_cons}.
The alternative energy proof parallels the entropy proof, illustrating
how these bounds allow finite-volume conservation laws to pass to
the thermodynamic limit.

\subsection{Surface energy and further properties of $H(\Psi|\Phi,t)$}
The next theorem gives conditions on $\Phi$ and $\Psi$ under which the
surface energy $W(\Psi|\Phi,t)$ exists and vanishes per unit volume
in the van~Hove limit. The proof is given in
Section~\ref{sec:surface_energies}.

\bet\label{thm:W_disapp}
Let $\Psi\in\leftindex_1\cB$ and $t\in\RR$.
\benr
\item If $\Phi\in\cBdiam{\gamma}$ with $\gamma>d$, then
$$
\Lim\frac1{|\Lambda|}
\left\|\tau_\Phi^{-t}(H_\Lambda(\Psi))-H_\Lambda(\Psi|\Phi,t)\right\|
=0.
$$
\item If $\Phi\in\cBdiam{\gamma}$ with $\gamma>2d$, then the
surface energy $W(\Psi|\Phi,t)$ of $H(\Psi|\Phi,t)$ exists and satisfies
$$
\Lim\frac1{|\Lambda|}\|W_\Lambda(\Psi|\Phi,t)\|=0.
$$
\eenr
In particular, under the assumptions of Part~\textup{(ii)}, if in addition
$\Psi\in\cBsd$, then the pair
$(\Phi,\Psi)$ satisfies~\cite[Condition~\textbf{(SD)}]{Jaksic2026a}.
\eet

The next result shows that the evolved state retains its equilibrium,
KMS, and Gibbs descriptions with respect to the transported
Hamiltonian family $H(\Psi|\Phi,t)$.

\bep\label{prop:state_properties}
Let $\Phi\in\cBdiam{\gamma}$ with $\gamma>2d$ and
$\Psi\in\leftindex_1\cB\cap\cBsd$. Let $t\in\RR$,
$\omega\in\cSeq(\Psi)$, and consider
$\omega_t=\omega\circ\tau^t_\Phi$. Then the following hold.
\benr
\item $\omega_t\in\cSeq(H(\Psi|\Phi,t))$.
\item $\omega_t$ is an $(\alpha_{H(\Psi|\Phi,t)},1)$-KMS state.
\item $\omega_t$ satisfies the Gibbs condition with respect to $H(\Psi|\Phi,t)$.
\eenr
\eep

\proof
\textbf{(i)} Since $\omega\in\cSeq(\Psi)$, one has
$s(\omega)=\omega(E(\Psi))+P(\Psi)$.
By Theorem~\ref{thm:entropy_energy_cons} and Proposition~\ref{prop:Ht_SDS}, we further
have $s(\omega)=s(\omega_t)$, $P_{H(\Psi|\Phi,t)}=P(\Psi)$ and
$E_{H(\Psi|\Phi,t)}=\tau^{-t}_\Phi(E(\Psi))$.
Hence, 
\[s(\omega_t)=\omega_t(E_{H(\Psi|\Phi,t)})+P_{H(\Psi|\Phi,t)},\]
 and so
$\omega_t\in\cSeq(H(\Psi|\Phi,t))$.

\medskip\noindent\textbf{(ii)}
Since $\Psi\in\cBsd$ and $\omega\in\cSeq(\Psi)$,
$\omega$ is an $(\tau_\Psi,1)$-KMS state
by~\cite[Theorems~A and~B]{Araki2003}.
For $A,B\in\fA$, set $A_t=\tau_\Phi^t(A)$ and $B_t=\tau_\Phi^t(B)$.
Proposition~\ref{prop:Ht_SDS}\textup{(iii)} gives
$$
\omega_t(A\alpha_{H(\Psi|\Phi,t)}^s(B))
=
\omega(A_t\tau_\Psi^s(B_t)),
$$
so $\omega_t$ is an $(\alpha_{H(\Psi|\Phi,t)},1)$-KMS state.

\medskip\noindent\textbf{(iii)}
Since the surface energies $W_\Lambda(\Psi|\Phi,t)$ are well-defined for
every $\Lambda$ (Proposition~\ref{prop:formula_W}),
\cite[Pro\-position~6.2.17]{Bratteli1981} implies that the
$(\alpha_{H(\Psi|\Phi,t)},1)$-KMS state $\omega_t$ satisfies the Gibbs
condition with respect to $H(\Psi|\Phi,t)$.
\qed

By conjugation, $\tau_\Phi^{-t}(\fAloc)$ is  a core for the generator
of the transported dynamics. The following theorem shows that $\fAloc$
itself is a core for this generator.

\bet\label{prop:core}
Suppose that $\Phi\in\cBdiam{\gamma}$ with $\gamma>2d$ and
$\Psi\in\cBsd$. For $t\in\RR$, let $\delta_t$ be the $\ast$-derivation
generating the dynamics
$$
\alpha_{H(\Psi|\Phi,t)}^s
=\tau_\Phi^{-t}\circ\tau_\Psi^s\circ\tau_\Phi^t.
$$
Then $\fAloc$ is a core for $\delta_t$ for every $t\in\RR$.
\eet
The proof is given in Section~\ref{sec:coreproof}. Together with
Theorem~\ref{thm:W_disapp}, this shows that if, in addition,
$\Psi\in\leftindex_1\cB$, then the pair $(\Phi,\Psi)$ satisfies
\cite[Condition~\textbf{(SD+)}]{Jaksic2026a}. All the properties entering
this condition hold for every $t\in\RR$.

\subsection{Constants of the motion}
\label{sec:Conservation}
We call $A\in\fAr$ a \textsl{constant of the motion} for $\tau$ if
$\nu\circ\tau^t(A)=\nu(A)$ for all $t\in\RR$ and $\nu\in\cSI$.
An interaction $\Psi$ is \textsl{$\tau_\Phi$-invariant} if its specific
energy $E(\Psi)$ is a constant of the motion for $\tau_\Phi$.
Here we work directly with observables and interactions. The formulation
modulo physical equivalence and the basic properties of these notions
are discussed in~\cite[Section~4]{Jaksic2026a}. Theorems~\ref{thm:W_disapp} and
\ref{thm:entropy_energy_cons}\textup{(i)}, together with
\cite[Theorems~4.5 and~4.8]{Jaksic2026a}, give the following theorem.

\bet\label{thm:comot}
Let $\Phi\in\cBdiam{\gamma}$ with $\gamma>2d$ and
$\Psi\in\leftindex_1\cB\cap\cBsd$. Then $\Psi$ is
$\tau_\Phi$-invariant iff
$\tau_\Phi^t\circ\tau_\Psi^s=\tau_\Psi^s\circ\tau_\Phi^t$
for all $s,t\in\RR$.
\eet

For the implication from invariance to commutation, the assumption
$\Psi\in\leftindex_1\cB$ can be omitted.

\subsection{Preservation of weak Gibbsianity}
\label{sec:WG}

Given two  interactions $\Psi$ and $\Phi$, consider again the family
$H(\Psi|\Phi,t)$ given by~\eqref{eq:HamHt}, and  let 
$\omega\in\cSwg(\Psi)$. We say that the weak Gibbsianity of $\omega$ is preserved 
by $\tau_\Phi$ at time $t$ if  $\omega_t=\omega\circ\tau^t_\Phi\in\cSwg(H(\Psi|\Phi,t))$.
The  next result gives conditions ensuring preservation of weak Gibbsianity.
It will be proved in Section~\ref{sec:proofwgPreserv} using a characterization of 
weak Gibbsianity derived in~\cite{Jaksic2023a}.

\bet\label{thm:wgPreserv}
Assume that $\Phi\in\cBf$.
\benr
\item If, for some $r>0$, $\|\Psi\|_r<r$, and $\omega\in\cSwg(\Psi)$, then
there exists $T>0$ such that $\omega_t\in\cSwg(H(\Psi|\Phi,t))$ for $|t|<T$.
\item If $d = 1$, $\Psi\in\cBf$, and $\omega\in\cSwg(\Psi)$, then 
$\omega_t\in\cSwg(H(\Psi|\Phi,t))$ for all $t\in\RR$.
\eenr
\eet
\ber
In case~\textup{(ii)}, the proof gives a stronger conclusion than
weak Gibbsianity: for every $t\in\RR$ there exists a constant $C_t\geq1$, independent of the finite interval
$\Lambda\subset\ZZ$, such that
$$
C_t^{-1}\omega_{H(\Psi|\Phi,t),\Lambda}
\leq
(\omega_t)_\Lambda
\leq
C_t\omega_{H(\Psi|\Phi,t),\Lambda},
$$
where
$$
\omega_{H(\Psi|\Phi,t),\Lambda}
=
\frac{\e^{-H_\Lambda(\Psi|\Phi,t)}}
{\tr(\e^{-H_\Lambda(\Psi|\Phi,t)})}.
$$
This follows from~\eqref{eq:boundsThm31} and the uniform bounds of
Theorem~\ref{thm:d1_extension}. In particular, for every $\nu\in\cSI$,
$$
\left|
S(\nu_\Lambda|(\omega_t)_\Lambda)
-
S(\nu_\Lambda|\omega_{H(\Psi|\Phi,t),\Lambda})
\right|
\leq
\log C_t.
$$
\eer


\subsection{Conservation of specific relative entropy}

The specific relative entropy of $\nu\in\cSI$ with respect to $\omega\in\cSI$ 
is defined as
$$
s(\nu|\omega)=\Llim\frac1{|\Lambda|}S(\nu_\Lambda|\omega_\Lambda),
$$
where $S(\nu_\Lambda|\omega_\Lambda)=\tr(\nu_\Lambda(\log\nu_\Lambda-\log\omega_\Lambda))$,
whenever the limit exists, in which case we have $s(\nu|\omega)\geq0$.

The conservation law for specific relative entropy follows from the
preservation of weak Gibbsianity together with the conservation of specific
entropy. Recall from~\cite{Jaksic2023a} that if $\omega\in\cSwg(\Psi)$, then
$(\omega,\Psi)$ is a regular pair, \ie $s(\nu|\omega)$ exists for all
$\nu\in\cSI$ and
$$
s(\nu|\omega)=-s(\nu)+\nu(E(\Psi))+P(\Psi).
$$

\bet
Assume that $(\omega,\Psi)$ is a regular pair and that either
\benr
\item $\|\Psi\|_r<r$ for some $r>0$, or
\item $d=1$ and $\Psi\in\cBf$.
\eenr
Let $\Phi\in\cBf$ and consider $\omega_t=\omega\circ\tau_\Phi^t$ for
$t\in\RR$. Then there exists $T>0$ such that, for every $\nu\in\cSI$ and
every $|t|<T$, the specific relative entropy $s(\nu_t|\omega_t)$, where
$\nu_t=\nu\circ\tau_\Phi^t$, exists and satisfies
$$
s(\nu_t|\omega_t)=s(\nu|\omega).
$$
In particular, $(\omega_t,H(\Psi|\Phi,t))$ is a regular pair for every
$|t|<T$. One can take $T=\infty$ in case~\textup{(ii)} or if $\omega$ is
$\tau_\Phi$-invariant.
\eet

\proof
Since $(\omega,\Psi)$ is a regular pair, $\omega\in\cSeq(\Psi)$.
Under either assumption~\textup{(i)} or~\textup{(ii)},
\cite[Theorem~2.11]{Jaksic2023a} gives
$\cSeq(\Psi)=\cSwg(\Psi)$, and hence $\omega\in\cSwg(\Psi)$.
With $T$ as in Theorem~\ref{thm:wgPreserv}, we have
$\omega_t\in\cSwg(H(\Psi|\Phi,t))$ for $|t|<T$, and one can take
$T=\infty$ in case~\textup{(ii)}. Since $\Phi\in\cBf$, the conservation
of specific entropy gives
$s(\nu_t)=s(\nu)$, while Proposition~\ref{prop:Ht_SDS} gives
$P_{H(\Psi|\Phi,t)}=P(\Psi)$ and
$\nu_t(E_{H(\Psi|\Phi,t)})=\nu(E(\Psi))$. Since $\omega_t\in\cSwg(H(\Psi|\Phi,t))$, the argument
of~\cite[Proposition~2.10]{Jaksic2023a}, applied to the family
$H(\Psi|\Phi,t)$, gives
\[
\begin{split}
s(\nu_t|\omega_t)
&=-s(\nu_t)+\nu_t(E_{H(\Psi|\Phi,t)})+P_{H(\Psi|\Phi,t)}\\
&=-s(\nu)+\nu(E(\Psi))+P(\Psi)\\
&=s(\nu|\omega),
\end{split}
\]
for $|t|<T$, and for all $t\in\RR$ in case~\textup{(ii)}.

Suppose now that $\omega$ is $\tau_\Phi$-invariant. Fix $t\in\RR$,
choose an integer $n\geq1$ such that $|t|/n<T$, and set $h=t/n$.
Since $\omega_h=\omega$, 
$s(\nu_{(k+1)h}|\omega)=s(\nu_{kh}|\omega)$ for
$k=0,\ldots,n-1$, and hence
$$
s(\nu_t|\omega)=s(\nu|\omega).
$$
Finally, for any $\nu\in\cSI$, regularity of $(\omega,\Psi)$ applied
to $\nu_{-t}$, together with the last identity, the conservation of
specific entropy, and
Proposition~\ref{prop:Ht_SDS}, gives
\[
\begin{split}
s(\nu|\omega)
&=s(\nu_{-t}|\omega)
=-s(\nu_{-t})+\nu_{-t}(E(\Psi))+P(\Psi)\\
&=-s(\nu)+\nu(E_{H(\Psi|\Phi,t)})+P_{H(\Psi|\Phi,t)}.
\end{split}
\]
Thus $(\omega,H(\Psi|\Phi,t))$ is regular for every $t\in\RR$.
\qed

\subsection{Hiai--Petz type bound}

Here we show that the vanishing of surface energies, Theorem~\ref{thm:W_disapp},
implies that the specific relative entropy, whenever it exists, does not exceed
its initial value along the trajectory. The following lemma is an adaptation
of~\cite[Lemma~2.3]{Hiai1993a} to our setting; see
also~\cite[Proposition~2.13]{Jaksic2023a}.
Recall that $\omega_{H(\Psi|\Phi,t),\Lambda}$ denotes the local Gibbs state
on $\Lambda$ for $H(\Psi|\Phi,t)$.

\bel\label{lem:Hiai}
Let $t\in\RR$. If $\omega\in\cSI$ satisfies the Gibbs condition for
$H(\Psi|\Phi,t)$, then
$$
S(\rho_\Lambda|\omega_\Lambda)
\leq
S(\rho_\Lambda|\omega_{H(\Psi|\Phi,t),\Lambda})
+2\|W_\Lambda(\Psi|\Phi,t)\|
$$
for any $\rho\in\cSI$ and $\Lambda\in\cF$.
\eel

\proof We follow~\cite[Lemma~2.3]{Hiai1993a}. $S(\argdot|\argdot)$ denotes Araki's relative entropy. 
 By assumption,
$$
\omega_{-W_\Lambda(\Psi|\Phi,t)}
=
\omega_{H(\Psi|\Phi,t),\Lambda}\otimes\nu
$$
for some state $\nu$ on $\fA_{\Lambda^c}$. Monotonicity under restriction
and the perturbation bound~\cite[Eq.~(2.3)]{Hiai1993a} give
\begin{align*}
S(\rho_\Lambda|\omega_\Lambda)
&\leq S(\rho_\Lambda\otimes\nu|\omega)\\
&\leq S(\rho_\Lambda\otimes\nu|
\omega_{-W_\Lambda(\Psi|\Phi,t)})
+2\|W_\Lambda(\Psi|\Phi,t)\|\\
&=S(\rho_\Lambda\otimes\nu|
\omega_{H(\Psi|\Phi,t),\Lambda}\otimes\nu)
+2\|W_\Lambda(\Psi|\Phi,t)\|\\
&=S(\rho_\Lambda|\omega_{H(\Psi|\Phi,t),\Lambda})
+2\|W_\Lambda(\Psi|\Phi,t)\|,
\end{align*}
as claimed.
\qed

\bep\label{prop:gibbs}
Let $\Phi\in\cBdiam{\gamma}$ with $\gamma>2d$ and
$\Psi\in\leftindex_1\cB\cap\cBsd$.
For $\nu\in\cSI$ and $\omega\in\cSeq(\Psi)$, set
$\nu_t=\nu\circ\tau_\Phi^t$ and $\omega_t=\omega\circ\tau_\Phi^t$.
Then, for every $t\in\RR$,
$$
\limsup_{\{\Lambda\}}\frac1{|\Lambda|}
S((\nu_t)_\Lambda|(\omega_t)_\Lambda)
\leq
-s(\nu)+\nu(E(\Psi))+P(\Psi).
$$
In particular, if $(\omega,\Psi)$ is a regular pair, the right-hand
side equals $s(\nu|\omega)$, and
$$
s(\nu_t|\omega_t)\leq s(\nu|\omega)
$$
whenever $s(\nu_t|\omega_t)$ exists.
\eep

\proof By Proposition~\ref{prop:state_properties}, $\omega_t$ satisfies
the Gibbs condition for $H(\Psi|\Phi,t)$. Since
Theorem~\ref{thm:W_disapp} gives
$\Llim|\Lambda|^{-1}\|W_\Lambda(\Psi|\Phi,t)\|=0$,
Lemma~\ref{lem:Hiai} yields
\begin{align*}
\limsup_{\{\Lambda\}}\frac1{|\Lambda|}
S((\nu_t)_\Lambda|(\omega_t)_\Lambda)
&\leq
\Llim\frac1{|\Lambda|}
S((\nu_t)_\Lambda|\omega_{H(\Psi|\Phi,t),\Lambda})
+2\Llim\frac1{|\Lambda|}\|W_\Lambda(\Psi|\Phi,t)\|\\[2pt]
&=
\Llim\frac1{|\Lambda|}
\left(
-S((\nu_t)_\Lambda)
+\tr((\nu_t)_\Lambda H_\Lambda(\Psi|\Phi,t))
+\log\tr\e^{-H_\Lambda(\Psi|\Phi,t)}
\right)\\[2pt]
&=
-s(\nu_t)
+\nu_t(E_{H(\Psi|\Phi,t)})
+P_{H(\Psi|\Phi,t)}.
\end{align*}
Theorem~\ref{thm:entropy_energy_cons} and
Proposition~\ref{prop:Ht_SDS} give
\beq
\limsup_{\{\Lambda\}}\frac1{|\Lambda|}
S((\nu_t)_\Lambda|(\omega_t)_\Lambda)
\leq
-s(\nu)+\nu(E(\Psi))+P(\Psi),
\label{eq:limsup_bound}
\eeq
as claimed. If $(\omega,\Psi)$ is regular, the right-hand side
of~\eqref{eq:limsup_bound} equals $s(\nu|\omega)$.
If $s(\nu_t|\omega_t)$ exists, it equals the limsup on the left.
\qed

\bec
Under the assumptions of Proposition~\ref{prop:gibbs}, suppose that
$(\omega,\Psi)$ is a regular pair and that $\omega$ is
$\tau_\Phi$-invariant. Then, for every $\nu\in\cSI$ and $t\in\RR$,
$$
s(\nu\circ\tau_\Phi^t|\omega)=s(\nu|\omega).
$$
\eec

\proof
Regularity of $(\omega,\Psi)$ implies that both relative entropies exist.
Applying Proposition~\ref{prop:gibbs} at time $t$ gives
$$
s(\nu\circ\tau_\Phi^t|\omega)\leq s(\nu|\omega).
$$
Applying the same proposition at time $-t$ to
$\nu\circ\tau_\Phi^t$ gives the reverse inequality.
\qed

\section{Proofs}
\label{sec:Proofs}

\subsection{Proof of Theorem~\ref{thm:bound_Lieb_Rob}}
\label{sec:proof_bound_Lieb_Rob}
We shall use the Lieb--Robinson bounds of~\cite{Nachtergaele2006,Nachtergaele2019},
formulated in terms of the following class of functions. A
non-increasing function $F\colon[0,\infty)\to(0,\infty)$ is called an
$F$-\textsl{function} when it satisfies
$$
\VERT F\VERT\coloneq\sum_{x\in\ZZ^d}F(|x|)<\infty
\quad\text{and}\quad
C_F\coloneq\sup_{x,y\in\ZZ^d}\sum_{z\in\ZZ^d}\frac{F(|x-z|)F(|z-y|)}{F(|x-y|)}<\infty.
$$
The corresponding norm on $\cB_\mathrm{f}$ is defined by
$$
\|\Phi\|_F\coloneq
\sup_{x\in\ZZ^d}\dfrac{1}{F(|x|)}
\sum_{X\supset\{0,x\}}\norm{\Phi(X)},
$$
and  $\cB_F$ denotes the associated Banach space of
interactions. If $F$ is an $F$-function and
$g\colon[0,\infty)\to[0,\infty)$ is non-negative, non-decreasing, and
sub-additive, then
$$
F_g\coloneq F\e^{-g}
$$
is again an $F$-function satisfying
$\VERT F_g\VERT\leq\VERT F\VERT$ and $C_{F_g}\leq C_F$; see
\cite[Section~8.2]{Nachtergaele2019}. We shall use the
bound~\cite[Eq.~(A21)]{Nachtergaele2019}
\beq
\sum_{(x,y)\in X\times Y}F_g(|x-y|)
\leq
\VERT F\VERT\min\{|X|,|Y|\}
\e^{-g(\dist(X,Y))},
\label{eq:bound_double_sum}
\eeq
which holds for all $X,Y\in\cF$.

For $\gamma>d$ and $0<\delta<\gamma-d$, the function
$$
F(u)\coloneq(1+u)^{-\gamma+\delta}
$$
is an $F$-function with $C_F\leq2^{\gamma-\delta}\VERT F\VERT$;
see~\cite[Section~8.1]{Nachtergaele2019}. Since 
$\{0,x\}\subset X \Rightarrow |x|\leq\diam(X)$, 
\beq
\|\Phi\|_F
\leq
\sum_{X\ni0}|X|(1+\diam(X))^{\gamma-\delta}\|\Phi(X)\|,
\label{mar-dot-1}
\eeq
and so  $\cBdiam{\gamma-\delta}\subset\cB_F$. With $g(u)\coloneq\delta\log(1+u)$, we have
\beq
F_g(u)=(1+u)^{-\gamma}.
\label{eq:Fg}
\eeq
The  estimate~\eqref{mar-dot-1} gives
$\|\Phi\|_{F_g}\leq\|\Phi\|_{\diam^\gamma}$, and hence
$\cBdiam{\gamma}\subset\cB_{F_g}$.

Specializing~\cite[Theorem~3.4(ii)]{Nachtergaele2019}
to a time-independent interaction gives the following proposition.
\bep\label{prop:LRboxes}
Let $\Phi\in\cBdiam{\gamma}$ for some $\gamma>d$. For any $t\in\RR$ and
$0<\delta<\gamma-d$, there exists a constant $c_{\Phi,\delta,t}$ such that
\beq
\left\|\tau_{\Phi,\Lambda}^t(A)
-\tau^t_{\Phi,\widetilde\Lambda}(A)\right\|
\leq
c_{\Phi,\delta,t}
\frac{|\Lambda'|}
{\left(1+\dist(\Lambda',\widetilde\Lambda\setminus\Lambda)\right)^\delta}
\|A\|
\label{mar-don}
\eeq
for every $\Lambda'\subset\Lambda\subset\widetilde\Lambda\in\cF$ and
$A\in\fA_{\Lambda'}$.
\eep

\proof
By the discussion above, $\Phi\in\cB_{F_g}$ with $F_g$ given by~\eqref{eq:Fg}. 
For $t\in\RR$ and $\Lambda'\subset\Lambda\subset\widetilde\Lambda\in\cF$,
\cite[Theorem~3.4(ii)]{Nachtergaele2019} gives
$$
\left\|\tau_{\Phi,\Lambda}^t(A)
-\tau^t_{\Phi,\widetilde\Lambda}(A)\right\|
\leq
\dfrac{2\norm{A}}{C_{F_g}}
\e^{2I_t(\Phi)}I_t(\Phi)
\sum_{(x,y)\in\Lambda'\times(\widetilde\Lambda\setminus\Lambda)}
F_g(|x-y|),
$$
where $I_t(\Phi)\coloneq C_{F_g}\norm{\Phi}_{F_g}|t|$. 
Using~\eqref{eq:bound_double_sum}, we obtain
$$
\left\|\tau_{\Phi,\Lambda}^t(A)
-\tau^t_{\Phi,\widetilde\Lambda}(A)\right\|
\leq
2\norm{A}\norm{\Phi}_{F_g}|t|
\e^{2C_{F_g}\norm{\Phi}_{F_g}|t|}
\VERT F\VERT|\Lambda'|
\e^{-g(\dist(\Lambda',\widetilde\Lambda\setminus\Lambda))}.
$$
This inequality and the bounds $C_{F_g}\leq C_F\leq2^\gamma\VERT F\VERT$ and
$\|\Phi\|_{F_g}\leq\|\Phi\|_{\diam^\gamma}$ yield~\eqref{mar-don} with
$$
c_{\Phi,\delta,t}
=
2\|\Phi\|_{\diam^\gamma}\VERT F\VERT |t|\,
\e^{2^{1+\gamma}\|\Phi\|_{\diam^\gamma}\VERT F\VERT |t|}.
$$
\qed

The inclusion $\cBdiam{\gamma}\subset\cBsd$ follows from the general theory
of $F$-interactions developed
in~\cite{Nachtergaele2006,Nachtergaele2019}. Letting
$\widetilde\Lambda$ exhaust $\ZZ^d$ in
Proposition~\ref{prop:LRboxes} gives the estimate in
Theorem~\ref{thm:bound_Lieb_Rob}.

\subsection{Proof of Theorem~\ref{thm:entropy_energy_cons}}
\label{sec:proof_entropy_energy_cons}

\subsubsection{Conservation of the specific entropy}
We follow the approach of~\cite{Wreszinski2019,Wreszinski2022},
using the entropy continuity bound of~\cite{Fannes1973} and
the Lieb--Robinson bounds of~\cite{Nachtergaele2019}.
\bel\label{lem:boxes} Fix $\gamma>2d$ and $d<\delta<\gamma-d$. Let $\{\Lambda\}$ be an
increasing family of cubes centered at the origin that exhausts $\ZZ^d$.
Then there exist concentric subcubes $\Lambda'\subset\Lambda$
 such that
\benr
\item $\Llim\frac{|\Lambda'|}{|\Lambda|}=1$.
\item $\Llim|\Lambda'|
\left(1+\dist(\Lambda',\Lambda^c)\right)^{-\delta}=0$.
\eenr
\eel
\proof
Write the cubes of the family as $\Lambda=\Lambda_L\coloneq\llbracket-L,L\rrbracket^d$ 
where $L\to\infty$. Choose $0<p<1-d/\delta$, set $r_L\coloneq\lfloor L^{1-p}\rfloor$, 
and define
$$
\Lambda'=\Lambda_L'
\coloneq
\llbracket-L+r_L,L-r_L\rrbracket^d.
$$

\textbf{(i)}
Since $r_L/L\to0$, we have
$$
\frac{|\Lambda_L'|}{|\Lambda_L|}
=
\left(
\frac{2(L-r_L)+1}{2L+1}
\right)^d
\rightarrow1.
$$

\noindent\textbf{(ii)}
Since
$$
\dist(\Lambda_L',\Lambda_L^c)=r_L+1,
$$
we have, for some constant $c>0$ and all sufficiently large $L$,
$$
|\Lambda_L'|
\left(1+\dist(\Lambda_L',\Lambda_L^c)\right)^{-\delta}
\leq
cL^{d-(1-p)\delta}.
$$
The exponent $d-(1-p)\delta$ is negative, so the right-hand side
tends to zero.
\qed

We denote the restriction of $\omega_t=\omega\circ\tau^t_\Phi$ to $\fA_\Lambda$ by
 $\omega_{t,\Lambda}$, and set
\[\omega_{\Lambda,t}=\omega_\Lambda\circ\tau^t_{\Phi,\Lambda}=\emitHLP \omega_\Lambda
\eitHLP.\]

The next proposition compares the entropies of $\omega_{t,\Lambda}$
and $\omega_{\Lambda,t}$.

\bep\label{prop:entropy}  Let $\Phi\in\cBdiam{\gamma}$ with $\gamma>2d$ and $\omega\in\cSI$. Let $\{\Lambda\}$ be an increasing family of cubes centered
at the origin that exhausts $\ZZ^d$. Then, for any $t$,  
$$
\Llim\frac1{|\Lambda|}
\left(S(\omega_{t,\Lambda})-S(\omega_{\Lambda,t})\right)=0.
$$
\eep
\proof  Fix $\delta\in(d,\gamma-d)$, choose $\{\Lambda'\}$ satisfying
Properties~(i)--(ii) of Lemma~\ref{lem:boxes}, and set
$\Lambda''=\Lambda\setminus\Lambda'$.  It suffices to show that for some constants $c_1, c_2$ and $c_3$,
\beq
\left|S(\omega_{t,\Lambda})-S(\omega_{\Lambda,t})\right|
\le c_1|\Lambda'|^2\left(1+\dist(\Lambda',\Lambda^c)\right)^{-\delta}
+c_2|\Lambda''|+c_3.
\label{eq:bound_h}
\eeq
Indeed,  after division by $|\Lambda|$,
the right-hand side tends to zero by Lemma~\ref{lem:boxes},
proving the result. We proceed to  prove~\eqref{eq:bound_h}.

Denote by $\omega_{\Lambda,t|\Lambda'}$ the restriction of
$\omega_{\Lambda,t}$ to $\fA_{\Lambda'}$.
The Araki--Lieb inequality~\cite{Araki1970} gives  
\begin{align*}
S(\omega_{t,\Lambda'})-S(\omega_{t,\Lambda''})
\le\left|S(\omega_{t,\Lambda'})-S(\omega_{t,\Lambda''})\right|
&\le S(\omega_{t,\Lambda})\le S(\omega_{t,\Lambda'})+S(\omega_{t,\Lambda''}),\\[4pt]
S(\omega_{\Lambda,t|\Lambda'})-S(\omega_{\Lambda,t|\Lambda''})
\le\left|S(\omega_{\Lambda,t|\Lambda'})-S(\omega_{\Lambda,t|\Lambda''})\right|
&\le S(\omega_{\Lambda,t})
\le S(\omega_{\Lambda,t|\Lambda'})+S(\omega_{\Lambda,t|\Lambda''}).
\end{align*}
It follows that
\begin{align*}
S(\omega_{t,\Lambda})-S(\omega_{\Lambda,t})&\le S(\omega_{t,\Lambda'})+S(\omega_{t,\Lambda''})-S(\omega_{\Lambda,t|\Lambda'})+S(\omega_{\Lambda,t|\Lambda''})\\
&\le\left|S(\omega_{t,\Lambda'})-S(\omega_{\Lambda,t|\Lambda'})\right|+S(\omega_{t,\Lambda''})+S(\omega_{\Lambda,t|\Lambda''}),
\end{align*}
and similarly
\begin{align*}
S(\omega_{\Lambda,t})-S(\omega_{t,\Lambda})&\le S(\omega_{\Lambda,t|\Lambda'})+S(\omega_{\Lambda,t|\Lambda''})-S(\omega_{t,\Lambda'})+S(\omega_{t,\Lambda''})\\
&\le\left|S(\omega_{t,\Lambda'})-S(\omega_{\Lambda,t|\Lambda'})\right|+S(\omega_{t,\Lambda''})+S(\omega_{\Lambda,t|\Lambda''}).
\end{align*}
Thus, 
\beq
\left|S(\omega_{t,\Lambda})-S(\omega_{\Lambda,t})\right|
\leq\left|S(\omega_{t,\Lambda'})-S(\omega_{\Lambda,t|\Lambda'})\right|+S(\omega_{t,\Lambda''})+S(\omega_{\Lambda,t|\Lambda''}),
\label{eq:box_outer00}
\eeq
where, with $c_2=2\log\dim\,\cH$,
\beq
S(\omega_{t,\Lambda''})+ S(\omega_{\Lambda,t|\Lambda''})\leq c_2|\Lambda''|.
\label{eq:box_outer}
\eeq
For these estimates, see also~\cite[Proof of Proposition~2.1]{Wreszinski2019}
and~\cite[Eq.~(56)]{Wreszinski2022}.

For the remaining term in~\eqref{eq:box_outer00}, we use
an argument of Fannes~\cite{Fannes1973}.
 Set $M=\dim\,\cH_{\Lambda'}=\e^{h|\Lambda'|}$, where $h=\log\dim\,\cH$, and denote by
$(\alpha_l)_{l\in\llbracket1,M\rrbracket}$ and $(\beta_l)_{l\in\llbracket1,M\rrbracket}$ 
the eigenvalues of $\omega_{t,\Lambda'}$ and
$\omega_{\Lambda,t|\Lambda'}$, respectively, counted with multiplicity
and arranged in increasing order.
Set $\epsilon_l=\beta_l-\alpha_l$, and note that, by~\cite[Theorem~6.11]{Kato1966},
\beq
a_{\Lambda,\Lambda'}\coloneq\sum_{l\in\llbracket1,M\rrbracket}|\epsilon_l|
\le\|\omega_{t,\Lambda'}-\omega_{\Lambda,t|\Lambda'}\|_1\le2,
\label{eq:abound}
\eeq
where $\|\argdot\|_1$ denotes the trace norm. Set $\fs(x)\coloneq-x\log x$ for $x\in(0,1]$, with $\fs(0)=0$.
By~\cite[Lemma~2]{Fannes1973}, for $x,y\in[0,1]$ we have
$$
|\fs(x)-\fs(y)|\leq\fs(|x-y|)+2|x-y|, 
$$
which implies that
\beq
\left|S(\omega_{t,\Lambda'})-S(\omega_{\Lambda,t|\Lambda'})\right|
\le\sum_{l\in\llbracket1,M\rrbracket}\fs(|\epsilon_l|)+4.
\label{eq:sum_eps0}
\eeq
If $a_{\Lambda,\Lambda'}=0$, then the left-hand side
of~\eqref{eq:sum_eps0} vanishes. Otherwise, invoking the concavity of the
logarithm, we get
\begin{align*}
\sum_{l\in\llbracket1,M\rrbracket}\fs(|\epsilon_l|)
&=a_{\Lambda,\Lambda'}
\sum_{\substack{l\in\llbracket1,M\rrbracket\\\epsilon_l\neq0}}
\frac{|\epsilon_l|}{a_{\Lambda,\Lambda'}}
\log\frac1{|\epsilon_l|}
\leq
a_{\Lambda,\Lambda'}
\log\sum_{\substack{l\in\llbracket1,M\rrbracket\\\epsilon_l\neq0}}
\frac1{a_{\Lambda,\Lambda'}}\\[4pt]
&\leq
a_{\Lambda,\Lambda'}(\log M-\log a_{\Lambda,\Lambda'})
\leq
ha_{\Lambda,\Lambda'}|\Lambda'|+\e^{-1},
\end{align*}
which, together with~\eqref{eq:sum_eps0}, yields
\beq
\left|S(\omega_{t,\Lambda'})-S(\omega_{\Lambda,t|\Lambda'})\right|
\leq ha_{\Lambda,\Lambda'}|\Lambda'|+c_3.
\label{eq:box_inner1}
\eeq
By duality, the first inequality in~\eqref{eq:abound} can be written as 
$$
a_{\Lambda,\Lambda'}\leq\sup_{\substack{A\in\fA_{\Lambda'}\\\|A\|\leq1}}
\left|\tr\left((\omega_{t,\Lambda'}-\omega_{\Lambda,t|\Lambda'})A\right)\right|,
$$
and since
$$
\left|\tr\left((\omega_{t,\Lambda'}-\omega_{\Lambda,t|\Lambda'})A\right)\right|
=\left|\omega\left(\tau_\Phi^t(A)-\tau^t_{\Phi,\Lambda}(A)\right)\right|
\leq\left\|\tau_\Phi^t(A)-\tau^t_{\Phi,\Lambda}(A)\right\|
$$
for $A\in\fA_{\Lambda'}$, it follows from Theorem~\ref{thm:bound_Lieb_Rob} that
$$
a_{\Lambda,\Lambda'}\leq\sup_{\substack{A\in\fA_{\Lambda'}\\\|A\|\leq1}}
\left\|\tau_\Phi^t(A)-\tau_{\Phi,\Lambda}^t(A)\right\|\leq c_t| \Lambda'|\left(1+\dist(\Lambda',\Lambda^c)\right)^{-\delta}.
$$
Combining the last inequality with~\eqref{eq:box_outer00}, \eqref{eq:box_outer}
and~\eqref{eq:box_inner1} proves~\eqref{eq:bound_h} with $c_1=hc_t$.
\qed

Since $\omega_{\Lambda,t}$ is unitarily equivalent to $\omega_\Lambda$,
$S(\omega_{\Lambda,t})=S(\omega_\Lambda)$. For
$\Phi\in\cBdiam{\gamma}$ with $\gamma>2d$,
Proposition~\ref{prop:entropy} gives
$$
s(\omega_t)
=\Llim\frac1{|\Lambda|}S(\omega_{t,\Lambda})
=\Llim\frac1{|\Lambda|}S(\omega_{\Lambda,t})
=\Llim\frac1{|\Lambda|}S(\omega_\Lambda)
=s(\omega),
$$
which proves Theorem~\ref{thm:entropy_energy_cons}\textup{(i)}.

\subsubsection{Conservation of the specific energy}
\bep\label{prop:energy_loc_glob}
Let $\Phi\in\cBdiam{\gamma}$ and assume either
$\gamma>2d$ and $\Psi\in\cBs$, or
$\gamma>d$ and $\Psi\in\leftindex_1\cB$.
Let $\{\Lambda\}$ be an increasing family of cubes centered at the origin
that exhausts $\ZZ^d$. Then, for any $\omega\in\cSI$ and $t$,
$$
\Llim\frac1{|\Lambda|}
\left|\left(\omega\circ\tau_\Phi^t-\omega\circ\tau^t_{\Phi,\Lambda}\right)
(H_\Lambda(\Psi))\right|
=
\Llim\frac1{|\Lambda|}
\left\|\left(\tau^t_\Phi-\tau^t_{\Phi,\Lambda}\right)
(H_\Lambda(\Psi))\right\|
=0.
$$
\eep

\proof
Since
$$
\left|\left(\omega\circ\tau_\Phi^t-\omega\circ\tau^t_{\Phi,\Lambda}\right)
(H_\Lambda(\Psi))\right|
\leq
\left\|\left(\tau^t_\Phi-\tau^t_{\Phi,\Lambda}\right)
(H_\Lambda(\Psi))\right\|,
$$
it suffices to prove that the second limit is zero.

Suppose first that $\gamma>d$ and $\Psi\in\leftindex_1\cB$. Fix
$0<\delta<\gamma-d$. Theorem~\ref{thm:bound_Lieb_Rob} gives
\begin{align*}
\left\|\left(\tau^t_\Phi-\tau^t_{\Phi,\Lambda}\right)
(H_\Lambda(\Psi))\right\|
&\leq
c_{\Phi,\delta,t}
\sum_{X\subset\Lambda}
\frac{|X|}{\left(1+\dist(X,\Lambda^c)\right)^\delta}
\|\Psi(X)\|\\
&\leq
c_{\Phi,\delta,t}\leftindex_1\|\Psi\|
\sum_{x\in\Lambda}
\frac1{\left(1+\dist(x,\Lambda^c)\right)^\delta}.
\end{align*}
For $r\in\NN$, set
\beq
\partial_r\Lambda
\coloneq
\left\{x\in\Lambda\mid\dist(x,\Lambda^c)\leq r\right\}.
\label{eq:deltarDef}
\eeq
Then
$$
\frac1{|\Lambda|}
\sum_{x\in\Lambda}
\frac1{\left(1+\dist(x,\Lambda^c)\right)^\delta}
\leq
\frac{|\partial_r\Lambda|}{|\Lambda|}
+
\frac1{(1+r)^\delta}.
$$
Taking the limit along the cubes and then letting $r\to\infty$ proves the
claim in this case.

It remains to consider $\gamma>2d$ and $\Psi\in\cBs$.
Fix $d<\delta<\gamma-d$ and choose the subcubes $\Lambda'$
provided by Lemma~\ref{lem:boxes}.  We claim
that, for some constants $c_1,c_2$,
\beq
\left\|\left(\tau^t_\Phi-\tau^t_{\Phi,\Lambda}\right)(H_\Lambda(\Psi))\right\|
\leq
c_1|\Lambda'|^2
\left(1+\dist(\Lambda',\Lambda^c)\right)^{-\delta}
+c_2|\Lambda\setminus\Lambda'|.
\label{eq:ineq1}
\eeq
The result  then follows from Lemma~\ref{lem:boxes}, as in the proof of
Proposition~\ref{prop:entropy}.

Set
$$
\tilde H_{\Lambda,\Lambda'}(\Psi)
\coloneq
H_\Lambda(\Psi)-H_{\Lambda'}(\Psi).
$$
Invoking Theorem~\ref{thm:bound_Lieb_Rob}, we obtain
\begin{align*}
\left\|\left(\tau^t_\Phi-\tau^t_{\Phi,\Lambda}\right)(H_\Lambda(\Psi))\right\|
&\leq
\left\|\left(\tau^t_\Phi-\tau^t_{\Phi,\Lambda}\right)
(H_{\Lambda'}(\Psi))\right\|
+
\left\|\tau^t_\Phi(\tilde H_{\Lambda,\Lambda'}(\Psi))\right\|
+
\left\|\tau^t_{\Phi,\Lambda}(\tilde H_{\Lambda,\Lambda'}(\Psi))\right\|\\[4pt]
&\leq
c_{\Phi,\delta,t}\|H_{\Lambda'}(\Psi)\||\Lambda'|
\left(1+\dist(\Lambda',\Lambda^c)\right)^{-\delta}
+2\|\tilde H_{\Lambda,\Lambda'}(\Psi)\|.
\end{align*}
We also have 
$$
\|H_{\Lambda'}(\Psi)\|
\leq
\sum_{X\cap\Lambda'\neq\emptyset}\|\Psi(X)\|
\leq
|\Lambda'|\sum_{X\ni0}\|\Psi(X)\|
\leq
|\Lambda'|\|\Psi\|_\mathrm{s},
$$
and similarly,
$$
\|\tilde H_{\Lambda,\Lambda'}(\Psi)\|
\leq
\sum_{X\cap(\Lambda\setminus\Lambda')\neq\emptyset}\|\Psi(X)\|
\leq
|\Lambda\setminus\Lambda'|\|\Psi\|_\mathrm{s}.
$$
Thus~\eqref{eq:ineq1} holds with
$c_1=c_{\Phi,\delta,t}\|\Psi\|_\mathrm{s}$ and
$c_2=2\|\Psi\|_\mathrm{s}$.
\qed

Let $\Phi\in\cBdiam{\gamma}$ with $\gamma>d$. Since
$\tau^t_{\Phi,\Lambda}(H_\Lambda(\Phi))=H_\Lambda(\Phi)$,
Eq.~\eqref{eq:spec_en_ham} and
Proposition~\ref{prop:energy_loc_glob}, applied with $\Psi=\Phi$, give
$$
(\omega\circ\tau_\Phi^t)(E(\Phi))-\omega(E(\Phi))
=
\Llim\frac1{|\Lambda|}
\left(\omega\circ\tau_\Phi^t-\omega\circ\tau^t_{\Phi,\Lambda}\right)
(H_\Lambda(\Phi))
=0.
$$
This proves Theorem~\ref{thm:entropy_energy_cons}\textup{(ii)}.

\subsection{Proof of Proposition~\ref{prop:Ht_SDS}}
\label{sec:proofPropHt_DS}

\noindent\textbf{(i)} Since  $\tr\,\e^{-H_\Lambda(\Psi|\Phi,t)}=\tr\,\e^{-H_\Lambda(\Psi)}$ 
for every $\Lambda\in\cF$, we have
$$
P(\Psi)=\Llim\frac1{|\Lambda|}\log\,\tr\e^{-H_\Lambda(\Psi)}
=\Llim\frac1{|\Lambda|}\log\,\tr\e^{-H_\Lambda(\Psi|\Phi,t)}= P_{H(\Psi|\Phi,t)}.
$$

\medskip\noindent\textbf{(ii)} 
Let $\omega\in\cSI$. Using Proposition~\ref{prop:energy_loc_glob} and
Eq.~\eqref{eq:spec_en_ham}, we obtain
\begin{align*}
\omega\circ\tau^{-t}_\Phi(E(\Psi))
&=\Llim\frac1{|\Lambda|}\omega\circ\tau^{-t}_\Phi(H_\Lambda(\Psi))\\
&=\Llim\frac1{|\Lambda|}\omega\circ\tau^{-t}_{\Phi,\Lambda}(H_\Lambda(\Psi))
=\Llim\frac1{|\Lambda|}\omega(H_\Lambda(\Psi|\Phi,t)),
\end{align*}
so $E_{H(\Psi|\Phi,t)}=\tau^{-t}_\Phi(E(\Psi))$ is a specific energy
observable of $H(\Psi|\Phi,t)$.

\medskip\noindent\textbf{(iii)}
The identity
$\e^{\i sH_\Lambda(\Psi|\Phi,t)}
=\e^{-\i tH_\Lambda(\Phi)}\e^{\i sH_\Lambda(\Psi)}
\e^{\i tH_\Lambda(\Phi)}$
gives
$$
\e^{\i sH_\Lambda(\Psi|\Phi,t)}A\e^{-\i sH_\Lambda(\Psi|\Phi,t)}
=\tau^{-t}_{\Phi,\Lambda}\circ\tau^s_{\Psi,\Lambda}
\circ\tau^t_{\Phi,\Lambda}(A).
$$
Since the finite-volume dynamics are isometric and converge in norm
on each observable, locally uniformly in time, the right-hand side
converges to $\tau_\Phi^{-t}\circ\tau_\Psi^s\circ\tau_\Phi^t(A)$,
uniformly for $(s,t)$ in compact subsets of $\RR^2$.

\subsection{Proof of Theorem~\ref{thm:W_disapp}}
\label{sec:surface_energies}
We will express the evolved surface energy as the sum of the evolved
original surface energy and two terms comparing the full dynamics
with its restrictions to $\Lambda$ and $\Lambda^c$.
Lieb--Robinson bounds will control these comparison terms and show
that they vanish per unit volume in the van~Hove limit.

For $K\subset\ZZ^d$,  $\tau_{\Phi|_K}$ denotes the dynamics
generated by the interaction $\Phi|_K$ defined by
$$
\Phi|_K(X)
=
\begin{cases}
\Phi(X)&\text{if }X\subset K;\\[4pt]
0&\text{otherwise.}  
\end{cases}
$$
Clearly $H_\Gamma(\Phi|_K)=H_{\Gamma\cap K}(\Phi)$, and hence 
$$
\tau_{\Phi|_K}^t(A)
=
\lim_{\{\Gamma\}}
\e^{\i tH_{\Gamma\cap K}(\Phi)}
A
\e^{-\i tH_{\Gamma\cap K}(\Phi)}
$$
for $A\in\fA$.
In particular, $\tau_{\Phi|_\Lambda}=\tau_{\Phi,\Lambda}$ for
$\Lambda\in\cF$.

We shall also use that these finite-volume approximations are independent
of the exhaustion. More precisely, if $\Phi\in\cBdiam{\gamma}$ with
$\gamma>d$, $K\subset\ZZ^d$, $A\in\fA$, and $t\in\RR$, then
$$
\tau_{\Phi|_K}^t(A)
=
\lim_\Gamma
\e^{\i tH_{\Gamma\cap K}(\Phi)}
A
\e^{-\i tH_{\Gamma\cap K}(\Phi)},
$$
where $\Gamma$ runs through $\cF$ directed by inclusion. For local $A$
this follows from Theorem~\ref{thm:bound_Lieb_Rob}; its proof applies to
$\Phi|_K$ with the same bound, since restriction only removes interaction
terms. The general case follows by density.

For $\delta>0$ and any non-empty $\Lambda\in\cF$, set
$$
b_\delta^{\rm in}(\Lambda)
\coloneq
\sum_{x\in\Lambda}
\frac1{\left(1+\dist(x,\Lambda^c)\right)^\delta}.
$$
If $\delta>d$, we also set
$$
b_\delta^{\rm out}(\Lambda)
\coloneq
\sum_{x\in\Lambda^c}
\frac1{\left(1+\dist(x,\Lambda)\right)^\delta}.
$$
One has
\beq
\begin{split}
\sum_{X\subset\Lambda}
\frac{|X|}{\left(1+\dist(X,\Lambda^c)\right)^\delta}\|\Psi(X)\|
&\leq
\leftindex_1\|\Psi\|\,b_\delta^{\rm in}(\Lambda),
\\[4pt]
\sum_{X\subset\Lambda^c}
\frac{|X|}{\left(1+\dist(X,\Lambda)\right)^\delta}\|\Psi(X)\|
&\leq
\leftindex_1\|\Psi\|\,b_\delta^{\rm out}(\Lambda),
\end{split}
\label{eq:est_surface_en2}
\eeq
where the first relation holds for every $\delta>0$, and the second one for
$\delta>d$. Indeed, for every
non-empty $X$ contained entirely in either $\Lambda$ or $\Lambda^c$, choose
$x_X\in X$ realizing the relevant distance:
$$
\begin{aligned}
\dist(x_X,\Lambda^c)&=\dist(X,\Lambda^c),
&&\text{if }X\subset\Lambda,\\
\dist(x_X,\Lambda)&=\dist(X,\Lambda),
&&\text{if }X\subset\Lambda^c.
\end{aligned}
$$
We group the interaction terms according to the chosen point $x_X$.
By translation invariance, each inner sum is bounded by
$$
\sum_{X\ni x}|X|\,\|\Psi(X)\|=\leftindex_1\|\Psi\|,
$$
which proves~\eqref{eq:est_surface_en2}.

We shall also use the van~Hove relations
\beq
\Lim\frac{b_\delta^{\rm in}(\Lambda)}{|\Lambda|}=0,
\qquad
\Lim\frac{b_\delta^{\rm out}(\Lambda)}{|\Lambda|}=0,
\label{eq:van_Hove_profiles}
\eeq
where the first relation holds for every $\delta>0$, and the second one for
$\delta>d$.  For the first relation, we use the bound
$$
\frac{b_\delta^{\rm in}(\Lambda)}{|\Lambda|}
\leq
\frac{|\partial_r\Lambda|}{|\Lambda|}
+
\frac1{(1+r)^\delta},
$$
valid for every $r\in\NN$, with $\partial_r\Lambda$ defined
in~\eqref{eq:deltarDef}. The conclusion follows by taking the
van~Hove limit and then letting $r\to\infty$.
For the second relation, set
$$
B_r
\coloneq
\left\{\left.z\in\ZZ^d\right| |z|\leq r\right\}.
$$
For each $x\in\Lambda^c$, choose $y_x\in\Lambda$ such that
$|x-y_x|=\dist(x,\Lambda)$. If $\dist(x,\Lambda)\leq r$, then
$y_x\in\partial_r\Lambda$. Moreover, the map
$x\mapsto(y_x,x-y_x)$ is injective. It follows that
$$
\frac{b_\delta^{\rm out}(\Lambda)}{|\Lambda|}
\leq
|B_r|\frac{|\partial_r\Lambda|}{|\Lambda|}
+
\sum_{\substack{z\in\ZZ^d\\|z|>r}}
\frac1{(1+|z|)^\delta}.
$$
Taking the van~Hove limit and then letting $r\to\infty$ proves the second
relation.

Finally, splitting the interactions contributing to $W_\Lambda(\Psi)$ according to their diameter gives
$$
\frac1{|\Lambda|}\|W_\Lambda(\Psi)\|
\leq
\frac{|\partial_r\Lambda|}{|\Lambda|}\|\Psi\|_\mathrm{s}
+
\sum_{\substack{X\ni0\\\diam(X)>r}}\|\Psi(X)\|.
$$
Since $\Psi\in\leftindex_1\cB\subset\cBs$, taking the van~Hove limit and then letting
$r\to\infty$ yields
\beq
\Lim\frac1{|\Lambda|}\|W_\Lambda(\Psi)\|=0.
\label{eq:surface_Psi_van_Hove}
\eeq

\bep\label{prop:formula_W}
Under the assumptions of Part~\textup{(ii)} of
Theorem~\ref{thm:W_disapp}, for each $\Lambda\in\cF$ the net
$$
\left(
W_{\Lambda,\widetilde\Lambda}(\Psi|\Phi,t)
\right)_{\Lambda\subset\widetilde\Lambda\in\cF}
$$
directed by inclusion converges in norm. In particular, the
surface energy
$W(\Psi|\Phi,t)=\left(W_\Lambda(\Psi|\Phi,t)\right)_{\Lambda\in\cF}$
of the Hamiltonian family $H(\Psi|\Phi,t)$ is well-defined and is given by
\beq
W_{\Lambda}(\Psi|\Phi,t)
=
\sum_{X\subset\Lambda}
\left(
\tau^{-t}_\Phi-\tau^{-t}_{\Phi|_\Lambda}
\right)(\Psi(X))
+
\sum_{X\subset\Lambda^c}
\left(
\tau^{-t}_\Phi-\tau^{-t}_{\Phi|_{\Lambda^c}}
\right)(\Psi(X))
+
\tau^{-t}_\Phi(W_\Lambda(\Psi)). 
\label{eq:surf_energy_formula}
\eeq
\eep

\proof
Fix $d<\delta<\gamma-d$, $\Lambda\in\cF$, and let
$\widetilde\Lambda$ run through the finite supersets of $\Lambda$, directed
by inclusion. By definition,
\beq
\begin{split}
W_{\Lambda,\widetilde\Lambda}(\Psi|\Phi,t)
&=
H_{\widetilde\Lambda}(\Psi|\Phi,t)
-
H_\Lambda(\Psi|\Phi,t)
-
H_{\widetilde\Lambda\setminus\Lambda}(\Psi|\Phi,t)
\\[8pt]
&=
\sum_{X\subset\widetilde\Lambda}
\tau^{-t}_{\Phi,\widetilde\Lambda}\left(\Psi(X)\right)
-
\sum_{X\subset\Lambda}
\tau^{-t}_{\Phi,\Lambda}\left(\Psi(X)\right)
-
\sum_{X\subset\widetilde\Lambda\setminus\Lambda}
\tau^{-t}_{\Phi,\widetilde\Lambda\setminus\Lambda}
\left(\Psi(X)\right)
\\[8pt]
&=
\sum_{X\subset\Lambda}
\left(
\tau^{-t}_{\Phi,\widetilde\Lambda}
-
\tau^{-t}_{\Phi,\Lambda}
\right)(\Psi(X))
+
\sum_{X\subset\widetilde\Lambda\setminus\Lambda}
\left(
\tau^{-t}_{\Phi,\widetilde\Lambda}
-
\tau^{-t}_{\Phi,\widetilde\Lambda\setminus\Lambda}
\right)(\Psi(X))
\\
&\qquad
+
\sum_{\mathclap{\twol{X\subset\widetilde\Lambda}
{X\cap\Lambda\neq\emptyset\neq
X\cap(\widetilde\Lambda\setminus\Lambda)}}}
\tau^{-t}_{\Phi,\widetilde\Lambda}(\Psi(X))
\\[4pt]
&=
\left(
\tau^{-t}_{\Phi,\widetilde\Lambda}
-
\tau^{-t}_{\Phi,\Lambda}
\right)(H_\Lambda(\Psi))
+
\left(
\tau^{-t}_{\Phi,\widetilde\Lambda}
-
\tau^{-t}_{\Phi,\widetilde\Lambda\setminus\Lambda}
\right)
(H_{\widetilde\Lambda\setminus\Lambda}(\Psi))
\\
&\qquad
+
\tau^{-t}_{\Phi,\widetilde\Lambda}
\left(W_{\Lambda,\widetilde\Lambda}(\Psi)\right).
\end{split}
\label{eq:est_surface_en0}
\eeq
Since $H_\Lambda(\Psi)\in\fA_\Lambda$ is fixed, the first term
after the last equality sign in~\eqref{eq:est_surface_en0} converges to
$$
\left(
\tau^{-t}_\Phi-\tau^{-t}_{\Phi,\Lambda}
\right)(H_\Lambda(\Psi)),
$$
which is the first term on the right-hand side
of~\eqref{eq:surf_energy_formula}. To deal with the second term, for $X\subset\Lambda^c$ set
$$
D_{\widetilde\Lambda}(X)
\coloneq
\begin{cases}
\left(
\tau^{-t}_{\Phi,\widetilde\Lambda}
-
\tau^{-t}_{\Phi,\widetilde\Lambda\setminus\Lambda}
\right)(\Psi(X))
&\text{if }X\subset\widetilde\Lambda\setminus\Lambda;\\[4pt]
0&\text{otherwise.}
\end{cases}
$$
Since $H_{\widetilde\Lambda}(\Phi|_{\Lambda^c})
=H_{\widetilde\Lambda\setminus\Lambda}(\Phi)$,
norm convergence of the finite-volume dynamics for $\Phi$ and
$\Phi|_{\Lambda^c}$ along the inclusion-directed net gives,
for every $X\subset\Lambda^c$,
$$
\lim_{\widetilde\Lambda}D_{\widetilde\Lambda}(X)
=
\left(
\tau^{-t}_{\Phi}
-
\tau^{-t}_{\Phi|_{\Lambda^c}}
\right)(\Psi(X)).
$$
By Proposition~\ref{prop:LRboxes}, there exists a constant
$c_{\Phi,\delta,t}$ such that
\beq
\|D_{\widetilde\Lambda}(X)\|
\leq
c_{\Phi,\delta,t}
\frac{|X|}
{\left(1+\dist(X,\Lambda)\right)^\delta}
\|\Psi(X)\|.
\label{eq:est_surface_en1}
\eeq
The right-hand side is summable over $X\subset\Lambda^c$
by~\eqref{eq:est_surface_en2}. Therefore, by dominated convergence,
\[
\lim_{\widetilde\Lambda}
\left(
\tau^{-t}_{\Phi,\widetilde\Lambda}
-
\tau^{-t}_{\Phi,\widetilde\Lambda\setminus\Lambda}
\right)
(H_{\widetilde\Lambda\setminus\Lambda}(\Psi))
=
\lim_{\widetilde\Lambda}
\sum_{X\subset\Lambda^c}D_{\widetilde\Lambda}(X)
=
\sum_{X\subset\Lambda^c}
\left(
\tau^{-t}_\Phi
-
\tau^{-t}_{\Phi|_{\Lambda^c}}
\right)(\Psi(X)).
\]
Finally we  consider
$\tau^{-t}_{\Phi,\widetilde\Lambda}
(W_{\Lambda,\widetilde\Lambda}(\Psi))$. Since
$\Psi\in\leftindex_1\cB\subset\cBs$, the series defining
$W_\Lambda(\Psi)$ is absolutely convergent, and so
$$
\lim_{\widetilde\Lambda}
\left\|
W_{\Lambda,\widetilde\Lambda}(\Psi)-W_\Lambda(\Psi)
\right\|
=0.
$$
Moreover,
\[
\left\|
\tau^{-t}_{\Phi,\widetilde\Lambda}
\left(W_{\Lambda,\widetilde\Lambda}(\Psi)\right)
-
\tau^{-t}_\Phi\left(W_\Lambda(\Psi)\right)
\right\|
\leq
\left\|
W_{\Lambda,\widetilde\Lambda}(\Psi)-W_\Lambda(\Psi)
\right\|
+
\left\|
\left(
\tau^{-t}_{\Phi,\widetilde\Lambda}
-
\tau^{-t}_\Phi
\right)
\left(W_\Lambda(\Psi)\right)
\right\|.
\]
The first term tends to zero by absolute convergence of the series
defining $W_\Lambda(\Psi)$. The second tends to zero because
$\tau^{-t}_{\Phi,\widetilde\Lambda}(W_\Lambda(\Psi))$
converges in norm to $\tau^{-t}_\Phi(W_\Lambda(\Psi))$
as $\widetilde\Lambda$ increases to $\ZZ^d$. Thus,
$$
\lim_{\widetilde\Lambda}
\tau^{-t}_{\Phi,\widetilde\Lambda}
\left(W_{\Lambda,\widetilde\Lambda}(\Psi)\right)
=
\tau^{-t}_\Phi\left(W_\Lambda(\Psi)\right).
$$
Taking the limit in~\eqref{eq:est_surface_en0} proves
\eqref{eq:surf_energy_formula}. \qed

\begin{remark}
The surface energy in Proposition~\ref{prop:formula_W} coincides with the
surface energy in the sense of Araki--Moriya~\cite[Equ.~(7.12)]{Araki2003}.
\end{remark}

We now conclude  the proof of Theorem~\ref{thm:W_disapp}.

\medskip\noindent\textbf{(i)}
Assume $\gamma>d$ and fix $0<\delta<\gamma-d$. By
Theorem~\ref{thm:bound_Lieb_Rob} and the first estimate
in~\eqref{eq:est_surface_en2},
\begin{align*}
\left\|
\tau_\Phi^{-t}(H_\Lambda(\Psi))
-
H_\Lambda(\Psi|\Phi,t)
\right\|
&=
\left\|
\left(
\tau_\Phi^{-t}
-
\tau_{\Phi,\Lambda}^{-t}
\right)(H_\Lambda(\Psi))
\right\|
\\
&\leq
c_{\Phi,\delta,t}
\sum_{X\subset\Lambda}
\frac{|X|}
{\left(1+\dist(X,\Lambda^c)\right)^\delta}
\|\Psi(X)\|
\\
&\leq
c_{\Phi,\delta,t}\leftindex_1\|\Psi\|\,
b_\delta^{\rm in}(\Lambda).
\end{align*}
After division by $|\Lambda|$, the result follows from the first relation
in~\eqref{eq:van_Hove_profiles}.

\medskip\noindent\textbf{(ii)}
Assume $\gamma>2d$ and fix $d<\delta<\gamma-d$. The existence of the
surface energy follows from
Proposition~\ref{prop:formula_W}. We consider the three terms on the
right-hand side of~\eqref{eq:surf_energy_formula}.

The first term vanishes after division by $|\Lambda|$ in the van~Hove
limit by Part~(i). For the second one, passing to the inclusion-directed limit
$\widetilde\Lambda\uparrow\ZZ^d$ in~\eqref{eq:est_surface_en1} and using
the second estimate in~\eqref{eq:est_surface_en2}, we obtain
\[
\left\|
\sum_{X\subset\Lambda^c}
\left(
\tau_\Phi^{-t}
-
\tau_{\Phi|_{\Lambda^c}}^{-t}
\right)(\Psi(X))
\right\|
\leq
c_{\Phi,\delta,t}
\sum_{X\subset\Lambda^c}
\frac{|X|}
{\left(1+\dist(X,\Lambda)\right)^\delta}
\|\Psi(X)\|
\leq
c_{\Phi,\delta,t}\leftindex_1\|\Psi\|\,
b_\delta^{\rm out}(\Lambda).
\]
The second relation in~\eqref{eq:van_Hove_profiles} then  yields
$$
\Lim\frac1{|\Lambda|}
\left\|
\sum_{X\subset\Lambda^c}
\left(
\tau_\Phi^{-t}
-
\tau_{\Phi|_{\Lambda^c}}^{-t}
\right)(\Psi(X))
\right\|
=0.
$$
Finally,
$$
\Lim\frac1{|\Lambda|}
\left\|
\tau_\Phi^{-t}(W_\Lambda(\Psi))
\right\|
=
\Lim\frac1{|\Lambda|}\|W_\Lambda(\Psi)\|
=0
$$
by~\eqref{eq:surface_Psi_van_Hove}. Combining these estimates with
\eqref{eq:surf_energy_formula} proves Part~(ii).

Theorem~\ref{thm:bound_Lieb_Rob} also gives $\Phi\in\cBsd$.
If, in addition, $\Psi\in\cBsd$, Parts~(i) and~(ii) show that
$(\Phi,\Psi)$ satisfies~\cite[Condition~\textbf{(SD)}]{Jaksic2026a}.

\subsection{Proof of Theorem~\ref{prop:core}}
\label{sec:coreproof}
Throughout this section we  assume that $\Phi\in\cBdiam{\gamma}$
with $\gamma>2d$ and $\Psi\in\cBsd$.  For fixed $t\in \RR$, 
$\delta_t$ denotes  the generator of
$\alpha^s=\tau_\Phi^{-t}\circ\tau_\Psi^s\circ\tau_\Phi^t$. Obviously, $\delta_t=\tau_\Phi^{-t}\circ\delta_\Psi\circ\tau_\Phi^t$ and 
$$
\Dom(\delta_t)=\tau_\Phi^{-t}(\Dom(\delta_\Psi)).
$$
Since $\fAloc$ is a core for $\delta_\Psi$, its image
$\tau_\Phi^{-t}(\fAloc)$ is a core for $\delta_t$, and to prove Theorem~\ref{prop:core} it suffices 
to approximate the elements of $\tau_\Phi^{-t}(\fAloc)$  by local observables in the
graph norm of $\delta_t$. To this end, we first verify that local observables
belong to $\Dom(\delta_t)$ and estimate the action of $\delta_t$
on them.

\bel\label{lem:delta_t}
Under the above assumptions, the following hold.
\benr
\item 
$\fA_\Lambda\subset\Dom(\delta_t)$ for any $\Lambda\in\cF$ . Moreover, there exists a constant $c_t$, depending
only on $\Phi$, $\Psi$, $\gamma$ and $t$, such that
\beq
\|\delta_t(A)\|\leq c_t|\Lambda|\|A\|
\label{eq:core_local_bound}
\eeq
for all $\Lambda\in\cF$ and $A\in\fA_\Lambda$.
\item If in addition $\Psi\in\leftindex_1\cB$, then 
$$
\delta_t(A)
=\i[H_\Lambda(\Psi|\Phi,t)+W_\Lambda(\Psi|\Phi,t),A]
=\lim_{\{\Lambda'\}}\i[H_{\Lambda'}(\Psi|\Phi,t),A]
$$
for any
$\Lambda\in\cF$ and $A\in\fA_\Lambda$.

\eenr
\eel

\proof
\textbf{(i)} Recall that  $F(u)=(1+u)^{-\gamma}$  is an $F$-function satisfying 
$\|\Phi\|_F\leq\|\Phi\|_{\diam^\gamma}$. The infinite-volume
Lieb--Robinson bound~\cite[Corollary~3.6(i)]{Nachtergaele2019} gives a
constant $L_t<\infty$, depending only on $\Phi$, $\gamma$ and $t$, such
that
\beq
\|[\tau_\Phi^t(C),D]\|
\leq L_t\|C\|\|D\|
\sum_{x\in X}\sum_{z\in Z}F(|x-z|)
\label{mar-bd}
\eeq
for disjoint $X,Z\in\cF$, $C\in\fA_X$ and $D\in\fA_Z$. Increasing $L_t$
to at least $2$ extends this bound to intersecting supports, since
$F(0)=1$ and $\|[\tau_\Phi^t(C),D]\|\leq2\|C\|\|D\|$.
Fix $A\in\fA_\Lambda$. Since $\Psi\in\cBsd\subset\cBs$, translation
invariance gives
$\sum_{Z\ni z}\|\Psi(Z)\|=\|\Psi\|_{\mathrm s}$ for every $z\in\ZZ^d$.
Summing  bound~\eqref{mar-bd} with $C=A$ and $D=\Psi(Z)$, we obtain
\beq
\begin{split}
	\sum_{Z\in\cF}\|[\Psi(Z),\tau_\Phi^t(A)]\|
	&\leq L_t\|A\|
	\sum_{x\in\Lambda}\sum_{z\in\ZZ^d}
	F(|x-z|)\sum_{Z\ni z}\|\Psi(Z)\|\\
	&=L_t\VERT F\VERT\|\Psi\|_{\mathrm s}|\Lambda|\|A\|.
\end{split}
\label{mar-tue}
\eeq
In particular,  setting $G=\tau_\Phi^t(A)$, the series
$$
S\coloneq\sum_{Z\in\cF}\i[\Psi(Z),G]
$$
converges  absolutely  in norm. We proceed to show that
$G\in\Dom(\delta_\Psi)$ and that $\delta_\Psi(G)=S$.

Let $\Gamma$ be a finite cube and 
$S_\Gamma=\i[H_\Gamma(\Psi),G]$. Absolute convergence implies that
$S_\Gamma$ converges in norm to $S$ as $\Gamma$ exhausts $\ZZ^d$. Write 
\beq
\tau_{\Psi,\Gamma}^s(G)-G
=\int_0^s\tau_{\Psi,\Gamma}^u(S_\Gamma)\d u.
\label{mar-phy}
\eeq
Obviously,  
$$
\|\tau_{\Psi,\Gamma}^u(S_\Gamma)-\tau_\Psi^u(S)\|
\leq\|S_\Gamma-S\|
+\|\tau_{\Psi,\Gamma}^u(S)-\tau_\Psi^u(S)\|,
$$
and since  $\Psi\in\cBsd$, the  right-hand side tends to zero uniformly for $u$ in compact intervals.
Passing to the limit in~\eqref{mar-phy}  gives
$$
\tau_\Psi^s(G)-G=\int_0^s\tau_\Psi^u(S)\d u.
$$
It follows that
$\lim_{s\to0}s^{-1}(\tau_\Psi^s(G)-G)=S$, and so 
$G\in\Dom(\delta_\Psi)$ and $\delta_\Psi(G)=S$, as claimed. Hence
$A\in\Dom(\delta_t)$ and $\delta_t(A)=\tau_\Phi^{-t}(S)$. Since $\|\delta_t(A)\|=\|S\|\leq
\sum_{Z\in\cF}\|[\Psi(Z),\tau_\Phi^t(A)]\|$,
the estimate~\eqref{mar-tue}  gives~\eqref{eq:core_local_bound} with
$c_t=L_t\VERT F\VERT\|\Psi\|_{\mathrm s}$.

\medskip\noindent\textbf{(ii)} Assume now that
$\Psi\in\leftindex_1\cB$ and write
$H_\Lambda(t)=H_\Lambda(\Psi|\Phi,t)$,
$W_\Lambda(t)=W_\Lambda(\Psi|\Phi,t)$ and
$W_{\Lambda,\Lambda'}(t)=W_{\Lambda,\Lambda'}(\Psi|\Phi,t)$.
For $\Lambda'\supset\Lambda$, the definition of the surface energy gives
$$
H_{\Lambda'}(t)
=H_\Lambda(t)+H_{\Lambda'\setminus\Lambda}(t)
+W_{\Lambda,\Lambda'}(t).
$$
$H_{\Lambda'\setminus\Lambda}(t)$  commutes with $A\in\fA_\Lambda$ and 
Proposition~\ref{prop:formula_W} gives norm convergence of
$W_{\Lambda,\Lambda'}(t)$ to $W_\Lambda(t)$. Hence
$$
\lim_{\{\Lambda'\}}\i[H_{\Lambda'}(t),A]
=\i[H_\Lambda(t)+W_\Lambda(t),A]\eqcolon D.
$$
Writing
$\alpha_{\Lambda'}^s=\tau_{\Phi,\Lambda'}^{-t}\circ
\tau_{\Psi,\Lambda'}^s\circ\tau_{\Phi,\Lambda'}^t$, we have
$$
\alpha_{\Lambda'}^s(A)-A
=\int_0^s\alpha_{\Lambda'}^u(\i[H_{\Lambda'}(t),A])\d u.
$$
Passing to the limit, 
$$
\alpha^s(A)-A=\int_0^s\alpha^u(D)\d u.
$$
Differentiating at $s=0$ gives  $\delta_t(A)=D$ and completes the proof of
the lemma.
\qed

Choose $p$ with $d<p<\gamma-d$,
which is possible since $\gamma>2d$. Let $X\in\cF$ be nonempty and
$A\in\fA_X$. Set $B=\tau_\Phi^{-t}(A)$ and, for $n\geq0$, let
$$
\Lambda_n\coloneq\{x\in\ZZ^d\mid\dist(x,X)\leq2^n\},
\qquad
B_n\coloneq\tau_{\Phi,\Lambda_n}^{-t}(A).
$$
Then $B_n\in\fA_{\Lambda_n}$,
$|\Lambda_n|\leq C_X2^{nd}$, and
$\dist(X,\Lambda_n^c)>2^n$. Applying
Theorem~\ref{thm:bound_Lieb_Rob} at time $-t$ with decay exponent $p$
gives 
\[
\|B-B_n\|\leq C2^{-np},
\]
where $C$ does not depend on $n$. 
It follows that $B_n$ converges to $B$ and that 
$$
\|B_n-B_{n-1}\|
\leq\|B_n-B\|+\|B-B_{n-1}\|
\leq C(1+2^p)2^{-np}.
$$
Since $\Lambda_{n-1}\subset\Lambda_n$, the difference $B_n-B_{n-1}$
belongs to $\fA_{\Lambda_n}$, and  Lemma~\ref{lem:delta_t}\textup{(i)} 
yields
\[
\|\delta_t(B_n)-\delta_t(B_{n-1})\|
\leq c_t|\Lambda_n|\|B_n-B_{n-1}\|
\leq c_tC_XC(1+2^p)2^{-n(p-d)}.
\]
Since $p>d$, the right-hand side is summable in $n$ and so 
$\delta_t(B_n)$ converges in norm. Since $\delta_t$ is closed, the limit of $\delta_t(B_n)$ is  $\delta_t(B)$. Hence $B_n$ converges to $B$ in the graph norm
of $\delta_t$. Finally, since $\tau_\Phi^{-t}(\fAloc)$ is a core for $\delta_t$, this shows that $\fAloc$ is also a core
for $\delta_t$. \qed

\subsection{Proof of Theorem~\ref{thm:wgPreserv}}
\label{sec:proofwgPreserv}
Throughout this section, $\Phi$, $\Psi$, and $\omega$ satisfy the
assumptions of either part of Theorem~\ref{thm:wgPreserv}.
In particular, Theorem~\ref{thm:W_disapp} applies. We write
$$
H(t)=H(\Psi|\Phi,t),\qquad
H_\Lambda(t)=H_\Lambda(\Psi|\Phi,t),\qquad
W_\Lambda(t)=W_\Lambda(\Psi|\Phi,t),\qquad
\alpha^s=\alpha^s_{H(t)}.
$$
Denoting by $(\omega_t)_{-W_\Lambda(t)}$ the perturbation of $\omega_t$ by the 
surface energy $-W_\Lambda(t)$, Proposition~\ref{prop:state_properties}\textup{(iii)}
gives
$$
\bigl((\omega_t)_{-W_\Lambda(t)}\bigr)_\Lambda=\omega_{H(t),\Lambda}
$$
for every $\Lambda\in\cF$. The argument of~\cite[Proposition~2.12]{Jaksic2023a} shows that
$\omega_t\in\cSwg(H(t))$ iff
\beq
\Llim\frac1{|\Lambda|}\log\inf_{\twol{A\in\fA_\Lambda}{A>0}}
\frac{\omega_t(A)}{{(\omega_t)}_{-W_\Lambda(t)}(A)}
=\Llim\frac1{|\Lambda|}\log\sup_{\twol{A\in\fA_\Lambda}{A>0}}
\frac{\omega_t(A)}{{(\omega_t)}_{-W_\Lambda(t)}(A)} 
= 0.
\label{eq:WG_char}
\eeq
We shall verify~\eqref{eq:WG_char} using the following criterion.
\bet\label{thm:WG} Let $\Phi$, $\Psi$, and $\omega$ satisfy the assumptions of either
part of Theorem~\ref{thm:wgPreserv}.
Fix $t$ and assume that there exists $a>\frac12$ such that,
for every cube $\Lambda$, the maps
$$
\RR\ni s\mapsto\alpha^s(W_\Lambda(t)),\qquad
\RR\ni s\mapsto\alpha^s_{-W_\Lambda(t)}(W_\Lambda(t))
$$
have analytic extensions to the strip $|\Im z|<a$, and
$$
\Llim\frac1{|\Lambda|}
\left(
\|\alpha^{\i/2}(W_\Lambda(t))\|
+
\|\alpha_{-W_\Lambda(t)}^{\i/2}(W_\Lambda(t))\|
\right)
=0.
$$
Then $\omega_t\in\cSwg(H(t))$.
\eet
\proof
By Proposition~\ref{prop:state_properties}\textup{(ii)}, $\omega_t$
is an $(\alpha,1)$-KMS state. Applying
\cite[Theorem~3.1]{Jaksic2023a} gives, for $A\in\fA_\Lambda$, $A>0$,
\beq
\e^{-\|W_\Lambda(t)\|-\|\alpha^{\i/2}(W_\Lambda(t))\|}
\leq
\frac{\omega_t(A)}{(\omega_t)_{-W_\Lambda(t)}(A)}
\leq
\e^{\|W_\Lambda(t)\|+
\|\alpha^{\i/2}_{-W_\Lambda(t)}(W_\Lambda(t))\|}.
\label{eq:boundsThm31}
\eeq
The assumptions and Theorem~\ref{thm:W_disapp} imply
\eqref{eq:WG_char}, proving the result.
\qed

To verify the assumptions of Theorem~\ref{thm:WG} and complete the proof
of Theorem~\ref{thm:wgPreserv}, we first consider arbitrary dimension
and sufficiently small $|t|$ in Section~\ref{sec:WG_everydim}.
The main ingredient is Proposition~\ref{prop:ext_Ruelle}, which
extends Ruelle's results~\cite[Section~7.6]{Ruelle1969}. 
In Section~\ref{sec:WG_1dim}, we treat dimension $d=1$ and arbitrary $t$,
using Proposition~\ref{prop:arakis_bound_dim1}, which adapts the
bound in~\cite{Araki1969}. 

We begin with some preliminary results.
 
\subsubsection{Preliminary results}
The  locally perturbed dynamics $\alpha_{-W_\Lambda(t)}$ can be factorized as follows. Set
$$
\alpha_K^s\coloneq\tau^{-t}_{\Phi|_K}\circ\tau^s_{\Psi|_K}\circ\tau^t_{\Phi|_K},
\qquad
\alpha_K^s(A)
=
\Llim
\e^{\i sH_{K\cap\Lambda}(t)}A\e^{-\i sH_{K\cap\Lambda}(t)}.
$$
\bel\label{lem:alphactor}
For any $\Lambda\in\cF$ and $s\in\RR$ one has 
\beq \alpha^s_{-W_{\Lambda}(t)}=\alpha^s_{\Lambda^c}\circ\alpha^s_\Lambda.
\label{mar-fac}
\eeq
\eel
\proof
We consider $s\geq0$; the case  $s<0$ is analogous. Let $(\tau_\iota)_{\iota\in I}$ be a net of $C^\ast$-dynamics on $\fA$
converging to a $C^\ast$-dynamics $\tau$ in norm on each observable,
uniformly for $s$ in compact subsets of $\RR$.
 If $(V_\iota)_{\iota\in I}$ is a
norm-convergent net in $\fAr$ with limit $V$, then
$$
\tau_V^s(A)=\lim_\iota(\tau_\iota)_{V_\iota}^s(A),
$$
where $\tau_V$ and $(\tau_\iota)_{V_\iota}$ denote the corresponding local
perturbations. This follows from the  Dyson expansion
$$
\tau_V^s(A)=\tau^s(A)+\sum_{n\ge1}\,
\int\limits_{0\le s_1\le\cdots\le s_n\le s}
\i[\tau^{s_1}(V),\i[\tau^{s_2}(V),\cdots\i[\tau^{s_n}(V),\tau^s(A)]\cdots]]
\d s_1\d s_2\cdots\d s_n,
$$
see, e.g., \cite[Proposition~5.4.1]{Bratteli1981}. On each compact time
interval, the corresponding integrands for $(\tau_\iota)_{V_\iota}$
converge uniformly to those above. The $n$-th integrated term is bounded by
$\|A\|(2M|s|)^n/n!$, where $M$ bounds $\|V_\iota\|$ on a tail of the net.
Uniform summability therefore proves the assertion. Moreover, if
$$
\tau_\iota^s(A)=\e^{\i sH_\iota}A\e^{-\i sH_\iota},
$$
then
$$
(\tau_\iota)_{V_\iota}^s(A)
=
\e^{\i s(H_\iota+V_\iota)}A\e^{-\i s(H_\iota+V_\iota)}.
$$
We fix $\Lambda\in\cF$ and $t\in\RR$.
By definition~\eqref{eq:surface_en_def}, for any $A$ in $\fA$ we have
$$
\alpha^s(A)=\lim_{\{\Lambda'\}}\e^{\i sH_{\Lambda'}(t)}A\e^{-\i sH_{\Lambda'}(t)}
=\lim_{\{\Lambda'\}}\e^{\i s(H_{\Lambda'\setminus\Lambda}(t)+H_\Lambda(t)+W_{\Lambda,\Lambda'}(t))}A
\e^{-\i s(H_{\Lambda'\setminus\Lambda}(t)+H_\Lambda(t)+W_{\Lambda,\Lambda'}(t))}.
$$
The convergence is uniform for $s$ in compact subsets of $\RR$,
by the finite-volume representation of $\alpha$ and the corresponding
convergence for $\tau_\Phi$ and $\tau_\Psi$.
 Thus, the previous argument yields
$$
\alpha^s_{-W_{\Lambda}(t)}(A)
=\lim_{\{\Lambda'\}}\e^{\i s(H_{\Lambda'\setminus\Lambda}(t)+H_\Lambda(t)+R_{\Lambda'})}A
\e^{-\i s(H_{\Lambda'\setminus\Lambda}(t)+H_\Lambda(t)+R_{\Lambda'})}
$$
where
$R_{\Lambda'}=W_{\Lambda,\Lambda'}(t)-W_\Lambda(t)$ converges to zero in
norm by Proposition~\ref{prop:formula_W}. It follows  that
\[
\begin{split}
\alpha^s_{-W_{\Lambda}(t)}(A)
&=\lim_{\{\Lambda'\}}\e^{\i s(H_{\Lambda'\setminus\Lambda}(t)+H_\Lambda(t))}A
\e^{-\i s(H_{\Lambda'\setminus\Lambda}(t)+H_\Lambda(t))}\\[1mm]
&=\lim_{\{\Lambda'\}}\e^{\i sH_{\Lambda^c\cap\Lambda'}(t)}\alpha_\Lambda^s(A)
\e^{-\i sH_{\Lambda^c\cap\Lambda'}(t)}\\[1mm]
&=\alpha_{\Lambda^c}^s\circ\alpha_\Lambda^s(A).
\end{split}
\]
\qed

The next lemma uses factorization~\eqref{mar-fac}  to establish analytic continuation
of $\alpha_{-W_\Lambda(t)}$ and bound its action on local observables.
\bel\label{lem:restricted_double_ineq} Let $t\in\RR$ and $\Lambda\in\cF$. Assume that there exists $a>0$
such that, for every $K\subset\ZZ^d$ and $A\in\fAloc$, the map
$$
\RR\ni s\mapsto\alpha_K^s(A)
$$
has an analytic extension to the strip
$S=\{z\in\CC\mid|\Im z|<a\}$.
Suppose also that there exists a non-decreasing function $\chi:\cF\to\RR_+$ such that
\beq
\|\alpha^z_K(A)\|\leq\chi(K\cap\supp A)\|A\| 
\label{mar-knee}
\eeq
for any $K\subset\ZZ^d$, $A\in\fAloc$ and $z\in S$. Then, for every
$A\in\fAloc$, the map
$$
\RR\ni s\mapsto\alpha^s_{-W_{\Lambda}(t)}(A)
$$
has an analytic extension to $S$, and for any $z\in S$ one has  
$$
\|\alpha^z_{-W_{\Lambda}(t)}(A)\|
\leq
\chi(\Lambda^c\cap\supp A)\chi(\Lambda\cap\supp A)\|A\|.
$$
\eel
\proof
For $A\in\fAloc$ and $z\in S$, set
$$
\beta^z(A)\coloneq\alpha^z_{\Lambda^c}\circ\alpha^z_\Lambda(A).
$$
Since
$$
\alpha^z_\Lambda(A)\in\fA_{\Lambda\cup\supp A},
$$
and $\fA_{\Lambda\cup\supp A}$ is finite-dimensional, an expansion of
$\alpha^z_\Lambda(A)$ in a fixed basis of this algebra shows that the map $S\ni z\mapsto\beta^z(A)$ 
is analytic. By Lemma~\ref{lem:alphactor},
$$
\beta^s(A)=\alpha^s_{-W_\Lambda(t)}(A)
$$
for $s\in\RR$. Thus, $\beta^z(A)$ is the analytic extension of
$\alpha^s_{-W_\Lambda(t)}(A)$ to $S$, and
$$
\alpha^z_{-W_\Lambda(t)}(A)
=
\alpha^z_{\Lambda^c}\circ\alpha^z_\Lambda(A).
$$
Moreover,
$$
\Lambda^c\cap\supp\alpha^z_\Lambda(A)
\subset
\Lambda^c\cap(\Lambda\cup\supp A)
=
\Lambda^c\cap\supp A.
$$
Applying  inequality~\eqref{mar-knee}  first with $K=\Lambda^c$ and then with
$K=\Lambda$, we obtain
\begin{align*}
\|\alpha^z_{-W_{\Lambda}(t)}(A)\|
&=\|\alpha^z_{\Lambda^c}\circ\alpha^z_\Lambda(A)\|\\
&\leq
\chi(\Lambda^c\cap\supp\alpha^z_\Lambda(A))
\|\alpha_\Lambda^z(A)\|\\
&\leq
\chi(\Lambda^c\cap\supp A)
\chi(\Lambda\cap\supp A)\|A\|,
\end{align*}
as claimed.
\qed

We next derive an expansion of $W_\Lambda(t)$ that will be used to
estimate $\|\alpha^z(W_\Lambda(t))\|$ and
$\|\alpha^z_{-W_\Lambda(t)}(W_\Lambda(t))\|$.
 For $\Phi\in\cBs$, let  $\delta_\Phi$ be the $\ast$-derivation defined on $\fAloc$ by
$$
\delta_\Phi(A)=\sum_{X\cap\supp A\neq\emptyset}\i[\Phi(X),A].
$$
We recall that $\delta_\Phi$ is closable, and  use the same notation for its closure. 
Recall also that $\Phi\in\cBsd$ if and only if
$(\Id\pm\delta_\Phi)\fAloc$ is dense in $\fA$;
see Section~\ref{sec:Inters}. For every $r>0$, one has $\cBf\subset\cB^r\subset\cBsd$.
For $\Phi\in\cB^r$,
$$
\|\delta_\Phi^n(A)\|
\leq n!\left(\frac{2\|\Phi\|_r}{r}\right)^n
\e^{r|\supp A|}\|A\|
$$
for $A\in\fAloc$ and $n\geq0$;
see~\cite[Theorem~6.2.4 and its proof]{Bratteli1981}.
In particular, the series
\beq
\tau_\Phi^z(A)=\sum_{n\geq0}\frac{z^n}{n!}\delta_\Phi^n(A)
\label{eq:Taylor}
\eeq
defines an analytic extension to the disc
$|z|<\Theta_\Phi\coloneq r/(2\|\Phi\|_r)$.
The group property extends it to the strip $|\Im z|<\Theta_\Phi$.

Let $D_\Phi(X)$ denote the $\ast$-derivation of $\fA$ defined by
$$
A\mapsto\i[\Phi(X),A].
$$
Given $\bY=(Y_1,\ldots,Y_n)\in\cF^n$, we further set $D_\Phi(\bY)=D_\Phi(Y_n)\cdots D_\Phi(Y_1)$. 

For $n\geq1$ and $X\in\cF$, we say that $\bY=(Y_1,\ldots,Y_n)\in\cF^n$ is an 
\emph{$(X,n)$-chain} whenever, for all $j\in\llbracket1,n\rrbracket$,
$$
Y_j\cap\left(X\cup\bigcup_{i=1}^{j-1}Y_i\right)\neq\emptyset.
$$
$\cC_n(X)$ denotes the set of all $(X,n)$-chains. The key observation, already made 
in~\cite[Lemma~7.6.1]{Ruelle1969}, is that for $A\in\fAloc$ and $\bY=(Y_1,\ldots,Y_n)\in\cF^n$, 
$$
D_\Phi(\bY)A\neq0\quad\Rightarrow\quad\bY\in\cC_n(\supp A),
$$
so that, for $n\ge1$,
$$
\delta_\Phi^n(A)=\sum_{\bY\in\cC_n(\supp A)}D_\Phi(\bY)A.
$$
Restricting the interaction to $K\subset\ZZ^d$ further reduces
the sum on the right-hand side.
Indeed, setting $\cC^K_n(X)=\{\bY\in\cC_n(X\cap K)\mid\cup_{i=1}^{n} Y_i \subset K\}$, 
we have
$$
D_{\Phi|_K}(\bY)A\neq0\quad\Rightarrow\quad\bY\in\cC^K_n(\supp A),
$$
and so 
\beq
\delta_{\Phi|_K}^n(A)=
\sum_{\bY\in\cF^n}
D_{\Phi|_K}(\bY)A
=\sum_{\bY\in\cC^K_n(\supp A)}D_{\Phi}(\bY)A.
\label{eq:need_chains3}
\eeq
We extend these definitions to the case $n=0$ by setting $\cC_0(X)=\cC_0^K(X)=\{\emptyset\}$ 
and $D_\Phi(\emptyset)=\Id$. Then, Relation~\eqref{eq:Taylor} becomes
\beq
\tau^z_{\Phi|_K}(A)=A+\sum_{n\ge1}\frac{z^n}{n!}\sum_{\bY\in\cC^K_n(\supp A)}D_\Phi(\bY)A
=\sum_{n\ge0}\frac{z^n}{n!}\sum_{\bY\in\cC^K_n(\supp A)}D_\Phi(\bY)A
\label{eq:evol_chains}
\eeq
for $|z|<\Theta_\Phi$. 

Finally, we introduce  the subset $\cC_n(X,\partial K)\subset\cC_n(X)$ consisting of $(X,n)$-chains $\bY$ which, 
when taken together with the starting set $X$, reach across the boundary of $K$, \ie satisfy
$$
K\cap\left(X\cup\bY\right)\neq\emptyset\qquad 
K^c\cap\left(X\cup\bY\right)\neq\emptyset,
$$
where
$$
X\cup\bY=\begin{cases}
X\cup\bigcup_{i=1}^nY_i&\text{if }\bY=(Y_1,\ldots,Y_n);\\
X&\text{if }n=0.
\end{cases}
$$
Set $\dist(X,\partial K)=\max(\dist(X,K),\dist(X,K^c))$ so that $\dist(X,\partial K)=0$
iff $X\cap K\neq\emptyset\neq X\cap K^c$ and hence
\beq
\cC_n(X,\partial K)=\begin{cases}
\cC_n(X)&\text{if }\dist(X,\partial K)=0,\\
\cC_n(X)\setminus\cC^K_n(X)&\text{if }X\subset K,\\
\cC_n(X)\setminus\cC^{K^c}_n(X)&\text{if }X\subset K^c.
\end{cases}
\label{eq:need_chains4}
\eeq
Note that, in particular,
$$
\cC_0(X,\partial K)=\begin{cases}
\{\emptyset\}&\text{if }\dist(X,\partial K)=0,\\
\emptyset&\text{otherwise.}
\end{cases}
$$
For $\Lambda\in\cF$, set
$$
\cF_\Lambda^\partial
\coloneq
\left\{X\in\cF\mid X\cap\Lambda\neq\emptyset\neq X\cap\Lambda^c\right\}.
$$

\bel\label{lem:WG_formula}  Let $\Lambda\in\cF$ and $\Phi\in\cBf$. Set
$R=\max\{\diam(X)\mid\Phi(X)\neq0\}$ if $\Phi\neq0$, and $R=0$ otherwise.
Then, for $|t|<\Theta_\Phi$, one has
\begin{align}
W_\Lambda(t)
&=W_\Lambda(\Psi)+\sum_{n\ge1}\frac{(-t)^n}{n!}
\sum_{X\in\cF,\bY\in\cC_n(X,\partial \Lambda)}D_\Phi(\bY)\Psi(X)
=\sum_{n\ge0}\frac{(-t)^n}{n!}
\sum_{X\in\cF,\bY\in\cC_n(X,\partial \Lambda)}D_\Phi(\bY)\Psi(X)\nonumber\\
&=W_\Lambda(\Psi)+\sum_{n\ge1}\frac{(-t)^n}{n!}
\sum_{\substack{X\in\cF\\
X\in\cF_\Lambda^\partial\ \text{or}\ X\cap\cT^\Lambda_n\neq\emptyset}}
\sum_{\bY\in\cC_n(X,\partial \Lambda)}D_\Phi(\bY)\Psi(X),
\label{eq:formula_Wt}
\end{align}
where $\cT^\Lambda_n=\{x\in\ZZ^d\mid\dist(\{x\},\partial\Lambda)\leq nR\}$.
\eel
\proof
The proof combines formula~\eqref{eq:surf_energy_formula} for $W_\Lambda(t)$ and the
expansion~\eqref{eq:evol_chains}. The first term on the right-hand side of~\eqref{eq:surf_energy_formula} 
gives, using~\eqref{eq:need_chains4},
\begin{align*}
\sum_{X\subset\Lambda}(\tau^{-t}_\Phi-\tau^{-t}_{\Phi|_\Lambda})(\Psi(X))
&=\sum_{X\subset\Lambda}\sum_{n\ge1}\frac{(-t)^n}{n!}\left(\sum_{\bY\in\cC_n(X)}D_\Phi(\bY)\Psi(X)
-\sum_{\bY\in\cC^\Lambda_n(X)}D_\Phi(\bY)\Psi(X)\right)\\[4pt]
&=\sum_{X\subset\Lambda }\sum_{n\ge1}\frac{(-t)^n}{n!}\sum_{\bY\in\cC_n(X,\partial\Lambda)}
D_\Phi(\bY)\Psi(X).
\end{align*}
For the second term, the same argument gives
$$
\sum_{X\subset\Lambda^c}(\tau^{-t}_\Phi-\tau^{-t}_{\Phi|_{\Lambda^c}})(\Psi(X))
=\sum_{X\subset\Lambda^c}\sum_{n\ge1}\frac{(-t)^n}{n!}
\sum_{\bY\in\cC_n(X,\partial\Lambda)}
D_\Phi(\bY)\Psi(X).
$$
For the third term, we use~\eqref{eq:surfacePsi} to obtain
$$
\tau^{-t}_\Phi(W_\Lambda(\Psi))=W_\Lambda(\Psi)+
\sum_{\twol{X\cap\Lambda\neq\emptyset}{X\cap\Lambda^c\neq\emptyset}}
\sum_{n\ge1}\frac{(-t)^n}{n!}
\sum_{\bY\in\cC_n(X,\partial\Lambda)}
D_\Phi(\bY)\Psi(X).
$$
These three formulas add to
$$
W_\Lambda(t) =W_\Lambda(\Psi)+
\sum_{X\in\cF}\sum_{n\ge1}\frac{(-t)^n}{n!}\sum_{\bY\in\cC_n(X,\partial\Lambda)}
D_\Phi(\bY)\Psi(X).
$$
We claim that this can be rewritten as~\eqref{eq:formula_Wt}. Indeed, let
$X\notin\cF_\Lambda^\partial$ and
$\bY=(Y_1,\ldots,Y_n)\in\cC_n(X,\partial\Lambda)$.
If $D_\Phi(\bY)\Psi(X)\neq0$, then $\diam(Y_i)\le R$ for every
$i\in\llbracket1,n\rrbracket$; the chain property  gives
$\dist(X,\partial\Lambda)\le nR$, and hence
$X\cap\cT^\Lambda_n\neq\emptyset$.
\qed

We shall use the following bound on the cardinality of $\cT^\Lambda_n$.
\bel\label{lem:size_Tk}
There exists a constant $c>0$ such that, for every cube $\Lambda\in\cF$
and every $n\geq1$, one has
$$
|\cT^\Lambda_n|\leq c|\Lambda|^{1-1/d}n^d.
$$
\eel
\proof We may assume that $\Lambda$ is centered at the origin.
If $2nR<|\Lambda|^{1/d}$, the set $\cT_n^\Lambda$ lies in a
boundary layer of thickness $nR$ on either side of $\Lambda$.
Hence, for some constant $c_1>0$,
$$
|\cT^\Lambda_n|\leq c_1|\Lambda|^{1-1/d}n^d.
$$
If $2nR\geq|\Lambda|^{1/d}$, then $\cT^\Lambda_n$ is contained in a cube centered at the origin
of side at most $|\Lambda|^{1/d}+2nR\leq4nR$, and hence
$$
|\cT^\Lambda_n|
\leq(4nR)^d
\leq c_2|\Lambda|^{1-1/d}n^d.
$$
Combining the two bounds yields the result.\qed

\subsubsection{Proof of Part~(i)}
\label{sec:WG_everydim}

In this section we assume without further mention that
\ben
\item $\Phi\in\cBf$ has range
$R\coloneq\max\{\diam(X)\mid\Phi(X)\neq0\}$, with the convention $R=0$
if $\Phi=0$;
\item $\|\Psi\|_r<r$ for some $r>0$.
\een
Choose $\rho,\eta>0$ such that
$$
\|\Psi\|_r<\rho,\qquad r_0\coloneq\rho+2\eta<r,
$$
and set
$$
\Theta_\Phi=\frac{\eta}{2\|\Phi\|_\eta},\qquad
\Theta_\Psi=\frac{\rho}{2\|\Psi\|_\rho}.
$$
Since $\|\Psi\|_\rho\leq\|\Psi\|_r<\rho$, one has $\Theta_\Psi>\frac12$.

The following lemma extends~\cite[Lemma~7.6.1]{Ruelle1969}
to interactions restricted to a subset $K\subset\ZZ^d$;
see also~\cite[Lemma~III.3.5]{Israel1979}.
The second inequality generalizes Ruelle's bound. The improvement
follows by restricting the sum over $\cC_n(X)$ to $\cC_n^K(X)$.

\bel\label{lem:Ruelle_chains}
For $K\subset\ZZ^d$, $X\in\cF$, $A\in\fAloc$, and $n\geq1$, one has
$$
\sum_{\bY\in\cC_n^K(X)}\prod_{i=1}^n\|\Psi(Y_i)\|\leq \e^{\rho|K\cap X|}\frac{n!}{(2\Theta_\Psi)^n},
$$
and
$$
\sum_{\bY\in\cF^n }\|D_{\Psi|_K}(\bY)A\|\leq\e^{\rho|K\cap\supp A|}\frac{n!}{\Theta_\Psi^n}\|A\|.
$$
\eel
\proof Set $X=\supp A$. By~\eqref{eq:need_chains3},
\beq
\sum_{\bY\in\cF^n}\|D_{\Psi|_K}(\bY)A\|
=\sum_{\bY\in\cC^K_n(X)}\|D_\Psi(\bY)A\|
\leq2^n\|A\|\sum_{\bY\in\cC_n^K(X)}
\prod_{i=1}^n\|\Psi(Y_i)\|.
\label{eq:Ruelle_est0}
\eeq
To bound the last sum, we follow~\cite[Eq.~(6.6)]{Ruelle1969}.
Let $\bY=(Y_1,\ldots,Y_n)\in\cC_n^K(X)$, set $Q_1\coloneq X\cap K$ and, for 
$i\in\llbracket2,n\rrbracket$, $Q_i\coloneq(X\cap K)\cup\bigcup_{\ell=1}^{i-1}Y_\ell$.
Since $|Q_i|\leq|X\cap K|+\sum_{\ell=1}^{i-1}(|Y_\ell|-1)$, we have
$$
\sum_{Y_i\cap Q_i\neq\emptyset}\|\Psi(Y_i)\|\leq|Q_i|\sum_{Z_i\ni0}\|\Psi(Z_i)\|
\leq\left(|X\cap K|+\sum_{\ell=1}^{i-1}(|Y_\ell|-1)\right)\sum_{k_i\ge0}
\sum_{\twol{Z_i\ni 0}{|Z_i|=k_i+1}}\|\Psi(Z_i)\|.
$$
Hence,
\begin{align*}
\sum_{\bY\in\cC_n^K(X)}\prod_{i=1}^n\|\Psi(Y_i)\|
&\leq
\sum_{Y_1\colon Y_1\cap Q_1\neq\emptyset}\cdots
\sum_{Y_n\colon Y_n\cap Q_n\neq\emptyset}
\prod_{i=1}^n\|\Psi(Y_i)\|\\
&\leq\sum_{(k_1,\ldots,k_n)\in\NN^n}\prod_{i=1}^n
\left((|X\cap K|+k_1+\ldots+k_{i-1})\sum_{\twol{Z_i\ni0}{|Z_i|=k_i+1}}\|\Psi(Z_i)\|\right).
\end{align*}
The inequality
$$
\prod_{i=1}^n(|X\cap K|+k_1+\ldots+k_{i-1})\leq
(|X\cap K|+k_1+\ldots+k_n)^n
\le\e^{\rho|X \cap K|}\frac{n!}{\rho^n} \prod_{i=1}^n \e^{\rho k_i}
$$
gives
$$
\sum_{\bY\in\cC_n^K(X)}\prod_{i=1}^n\|\Psi(Y_i)\|
\leq\frac{n!}{\rho^n}\e^{\rho|X\cap K|}
\left(\sum_{Z\ni0}\e^{\rho(|Z|-1)}\|\Psi(Z)\|\right)^n
\leq n!\,\e^{\rho|X\cap K|}\left(\frac{\|\Psi\|_\rho}{\rho}\right)^n. 
$$
This proves the first bound. The second follows
from~\eqref{eq:Ruelle_est0}. \qed

Applying Lemma~\ref{lem:Ruelle_chains} to the Taylor
expansion~\eqref{eq:evol_chains} gives the following extension
of~\cite[Theorem~2.3]{Jaksic2023a}.
\bep\label{prop:ext_Ruelle} 
For all $K\subset\ZZ^d$ and $A\in\fAloc$, the map
$$
\RR\ni s\mapsto\tau_{\Psi|_K}^s(A)\in\fA
$$
has an analytic extension to the strip $|\Im z|<\Theta_\Psi$, and for any $z$ in this strip
$$
\|\tau_{\Psi|_K}^z(A)\|\leq\frac{\e^{\rho|K\cap\supp A|}}{1-|\Im z|/\Theta_\Psi}\|A\|.
$$
\eep

The next proposition shows that $\alpha_K$ satisfies the assumptions
of Lemma~\ref{lem:restricted_double_ineq} for sufficiently small $|t|$.
\bep\label{prop:alphaK_exten} Fix $0<\theta<\Theta_\Psi$. Let
$\kappa_R\coloneq|\{x\in\ZZ^d\mid |x|\leq R\}|$
be the number of lattice sites in the ball of radius $R$, and set
$$
T_\theta\coloneq
\e^{-\rho\kappa_R}\Theta_\Phi\left(1-\theta/\Theta_\Psi\right).
$$
Then, for all $K\subset\ZZ^d$, $|t|<T_\theta$ and $A\in\fAloc$, the map 
$$
\RR\ni s\mapsto\alpha_K^s(A)\in\fA
$$
has an analytic extension to the strip $|\Im z|<\theta$, and for any $z$ in this strip
$$
\|\alpha^z_K(A)\|\leq
\frac{\e^{(\rho+\eta)|\supp A\cap K|}}
{\left(1-\theta/\Theta_\Psi\right)\left(1-|t|/T_\theta\right)}\|A\|.
$$
\eep
\proof
Set $C_{\Psi,\theta}=(1-\theta/\Theta_\Psi)^{-1}$. Let $A\in\fAloc$ and set $X=\supp A$. 
For real $z$, Eq.~\eqref{eq:evol_chains} gives
$$
\tau^z_{\Psi|_K}\circ\tau^t_{\Phi|_K}(A)
=\tau^z_{\Psi|_K}(A)+\sum_{n\ge1}\frac{t^n}{n!}\sum_{\bY\in\cC^K_n(X)}\tau^z_{\Psi|_K}(D_\Phi(\bY)A).
$$
For complex $z$ with $|\Im z|<\theta$, we estimate the series on
the right-hand side. The bounds below will establish its locally
uniform convergence and justify the notation
$\tau^z_{\Psi|_K}\circ\tau^t_{\Phi|_K}(A)$.
After applying the isometry $\tau^{-t}_{\Phi|_K}$, we obtain
\beq
\begin{split}
\|\alpha^z_K(A)\|=\|\tau^z_{\Psi|_K}\circ\tau^t_{\Phi|_K}(A)\|
&\le\|\tau^z_{\Psi|_K}(A)\|+\sum_{n\ge1}\frac{|t|^n}{n!}\sum_{\bY\in\cC_n^K(X)}\|\tau_{\Psi|_K}^z(D_\Phi(\bY)A)\|\\ 
&\leq\left(1+\sum_{n\ge1}\frac{(2|t|)^n}{n!}\sum_{\bY\in\cC_n^K(X)}
\prod_{i=1}^n\|\tau_{\Psi|_K}^z(\Phi(Y_i))\|\right)\| \tau_{\Psi|_K}^z(A)\|.
\end{split}
\label{eq:WG_bounds}
\eeq
Proposition~\ref{prop:ext_Ruelle} gives
$$
\|\tau^z_{\Psi|_K}(A)\|\leq C_{\Psi,\theta}\e^{\rho|K\cap X|}\|A\|,
$$
and 
$$
\|\tau^z_{\Psi|_K}(\Phi(Y))\|\leq C_{\Psi,\theta}\e^{\rho\kappa_R}\|\Phi(Y)\|
$$
for every $Y\in\cF$. Moreover, Lemma~\ref{lem:Ruelle_chains}, applied to
$\Phi$ with the parameter $\eta$, gives
$$
\sum_{\bY\in\cC_n^K(X)}\prod_{i=1}^n\|\Phi(Y_i)\|
\le\e^{\eta|K\cap X|}\frac{n!}{(2\Theta_\Phi)^n}
$$
for every $n\ge1$. Substituting  these bounds into Eq.~\eqref{eq:WG_bounds}, we obtain
\begin{align*}
\|\alpha^z_K(A)\|
&\leq C_{\Psi,\theta}\e^{(\rho+\eta)|K\cap X|}\|A\|
\sum_{n\ge0}\left(\tfrac{2}{\eta}\|\Phi\|_\eta\e^{\rho\kappa_R}C_{\Psi,\theta}|t|\right)^n\\
&=\frac{\e^{(\rho+\eta)|K\cap X|}}
{1-(\theta/\Theta_\Psi+\e^{\rho\kappa_R}|t|/\Theta_\Phi)}\|A\|\\
&=\frac{\e^{(\rho+\eta)|K\cap\supp A|}}
{\left(1-\theta/\Theta_\Psi\right)\left(1-|t|/T_\theta\right)}\|A\|.         
\end{align*}
These bounds establish locally uniform convergence of the series
on $|\Im z|<\theta$. Its sum agrees with
$\tau^z_{\Psi|_K}\circ\tau^t_{\Phi|_K}(A)$ for real $z$.
Applying $\tau^{-t}_{\Phi|_K}$ therefore gives the required analytic
extension of $\alpha_K^s(A)$, with the stated bound.
\qed

The next theorem allows us to apply Theorem~\ref{thm:WG}
for sufficiently small $|t|$ and complete the proof of
Theorem~\ref{thm:wgPreserv}\textup{(i)}.

\bet\label{thm:alphaW_final}
For any $0<\theta<\Theta_\Psi$ there exists $\bar T_\theta>0$ such that, for any $|t|<\bar T_\theta$ and
any cube $\Lambda\in\cF$, the maps
$$
\RR\ni s\mapsto\alpha^s(W_\Lambda(t)),\qquad \RR\ni s\mapsto\alpha_{-W_{\Lambda}(t)}^s(W_\Lambda(t)),
$$ 
have analytic extensions to the strip $|\Im z|<\theta$. Moreover, for $z$ in this strip,
$$
\Llim\frac1{|\Lambda|}\left(\|\alpha^{z}(W_\Lambda(t))\|+\|\alpha_{-W_{\Lambda}(t)}^z(W_\Lambda(t))\|\right)=0.
$$
\eet

\proof Fix $0<\theta<\Theta_\Psi$ and $0<\bar t<T_\theta$.
Suppose that $|\Im z|<\theta$ and $|t|<\bar t$.
We first apply $\alpha^z$ termwise to the expansion in
Lemma~\ref{lem:WG_formula}, expanding $W_\Lambda(\Psi)$ into
its interaction terms. The estimates below establish locally
uniform convergence and justify denoting the resulting sum by
$\alpha^z(W_\Lambda(t))$. We obtain
\begin{equation}
\begin{split}
&\|\alpha^z(W_\Lambda(t))\|
\le\|\alpha^z(W_\Lambda(\Psi))\|+\sum_{n\ge1}\frac{\bar t^n}{n!}
\sum_{\substack{X\in\cF\\
X\in\cF_\Lambda^\partial\ \text{or}\ X\cap\cT^\Lambda_n\neq\emptyset}}
\sum_{\bY\in\cC_n(X,\partial\Lambda)}\|\alpha^z(D_\Phi(\bY)\Psi(X))\|\\
&\le\sum_{X\in\cF_\Lambda^\partial}\|\alpha^z(\Psi(X))\|
+\sum_{n\ge1}\frac{(2\bar t)^n}{n!}
\sum_{\substack{X\in\cF\\
X\in\cF_\Lambda^\partial\ \text{or}\ X\cap\cT^\Lambda_n\neq\emptyset}}
\|\alpha^z(\Psi(X))\|\sum_{\bY\in\cC_n(X,\partial\Lambda)}
\prod_{i=1}^n\|\alpha^z(\Phi(Y_i))\|.
\end{split}
\label{eq:salome}
\end{equation}
For $X,Y\in\cF$, Proposition~\ref{prop:alphaK_exten},
applied with $K=\ZZ^d$, gives
\beq
\|\alpha^z(\Psi(X))\|\leq\tilde C\e^{(\rho+\eta)|X|}\|\Psi(X)\|,\qquad 
\|\alpha^z(\Phi(Y))\|\leq\tilde C\e^{(\rho+\eta)\kappa_R}\|\Phi(Y)\|,
\label{eq:clouds}
\eeq
where $\tilde C\coloneq(1-\theta/\Theta_\Psi)^{-1}(1-\bar t/T_\theta)^{-1}$. 
Since $\rho+\eta<r$, \eqref{eq:clouds} gives
$$
\frac1{|\Lambda|}\sum_{\twol{X\cap\Lambda\neq\emptyset}{X\cap\Lambda^c\neq\emptyset}}\|\alpha^z(\Psi(X))\|
\le\frac{\tilde C}{|\Lambda|}
\sum_{x\in\Lambda}\sum_{\twol{X\ni0}{X\cap(\Lambda^c+x)\neq\emptyset}}
\e^{(\rho+\eta)|X|}\|\Psi(X)\|\rightarrow0,
$$
along centered cubes. The series over $n\geq1$ in~\eqref{eq:salome} is bounded by
$$
M_\Lambda\coloneq
\tilde C\sum_{n\ge1}\frac{(2\tilde C\e^{(\rho+\eta)\kappa_R}\bar t)^n}{n!}
\sum_{\substack{X\in\cF\\
X\in\cF_\Lambda^\partial\ \text{or}\ X\cap\cT^\Lambda_n\neq\emptyset}}
\e^{(\rho+\eta)|X|}\|\Psi(X)\|
\sum_{\bY\in\cC_n(X,\partial\Lambda)}\prod_{i=1}^n\|\Phi(Y_i)\|.
$$
Applying Lemma~\ref{lem:Ruelle_chains} to $\Phi$ with the parameter $\eta$, we get
$$
\sum_{\bY\in\cC_n(X,\partial\Lambda)}\prod_{i=1}^n\|\Phi(Y_i)\|
\le\e^{\eta|X|}\frac{n!}{(2\Theta_\Phi)^n}
$$
for every $X\in\cF$ and $n\in\NN$. Lemma~\ref{lem:size_Tk} further gives  
$$
\sum_{X\cap\cT_n^\Lambda\neq\emptyset}\e^{r_0|X|}\|\Psi(X)\|
\le\e^{r_0}\sum_{x\in\cT_n^\Lambda}\sum_{X\ni x}\e^{r_0(|X|-1)}\|\Psi(X)\|
\leq c\e^{r_0}\|\Psi\|_{r_0}|\Lambda|^{1-1/d}n^d.
$$
Set
$$
b_\Lambda\coloneq
\frac1{|\Lambda|}\sum_{X\in\cF_\Lambda^\partial}\e^{r_0|X|}\|\Psi(X)\|.
$$
Since $r_0<r$,
$$
b_\Lambda
\leq
\frac1{|\Lambda|}
\sum_{x\in\Lambda}
\sum_{\twol{X\ni0}{X\cap(\Lambda^c+x)\neq\emptyset}}
\e^{r_0|X|}\|\Psi(X)\|
\rightarrow0
$$
along centered cubes. Consequently,
$$
\frac1{|\Lambda|}M_\Lambda\leq
\tilde C\sum_{n\ge1}
\left(\frac{\tilde C\e^{(\rho+\eta)\kappa_R}\bar t}{\Theta_\Phi}\right)^n
\left(c\e^{r_0}\|\Psi\|_{r_0}|\Lambda|^{-1/d}n^d+b_\Lambda\right)
\rightarrow0
$$
along centered cubes, provided that
$$
\bar t<T_1\coloneq\frac{T_\theta}{1+\e^{\eta\kappa_R}},
$$
which ensures convergence of the series.

For $\alpha_{-W_\Lambda(t)}$, the same argument gives
\beq
\|\alpha_{-W_{\Lambda}(t)}^z(W_\Lambda(t))\|
\le\sum_{X\in\cF_\Lambda^\partial}
\|\alpha_{-W_{\Lambda}(t)}^z(\Psi(X))\|
+M_\Lambda',
\label{eq:rain}
\eeq
where
$$
M_\Lambda'=\sum_{n\ge1}\frac{(2\bar t)^n}{n!}
\sum_{\substack{X\in\cF\\
X\in\cF_\Lambda^\partial\ \text{or}\ X\cap\cT^\Lambda_n\neq\emptyset}}
\|\alpha_{-W_{\Lambda}(t)}^z(\Psi(X))\|
\sum_{\bY\in\cC_n(X,\partial\Lambda)}
\prod_{i=1}^n\|\alpha_{-W_{\Lambda}(t)}^z(\Phi(Y_i))\|.
$$
By Proposition~\ref{prop:alphaK_exten} and Lemma~\ref{lem:restricted_double_ineq}, the 
estimates~\eqref{eq:clouds} are replaced by
$$
\|\alpha^z_{-W_{\Lambda}(t)}(\Psi(X))\|
\leq\tilde C^2\e^{(\rho+\eta)|X|}\|\Psi(X)\|,\qquad
\|\alpha^z_{-W_{\Lambda}(t)}(\Phi(Y))\|
\leq\tilde C^2\e^{(\rho+\eta)\kappa_R}\|\Phi(Y)\|.
$$
Proceeding as above, the first term on the right-hand side of~\eqref{eq:rain} is
$o(|\Lambda|)$ along the centered cubes. For the second term, we get the estimate
$$
\frac1{|\Lambda|}M_\Lambda'\leq
\tilde C^2\sum_{n\ge1}
\left(\frac{\tilde C^2\e^{(\rho+\eta)\kappa_R}\bar t}{\Theta_\Phi}\right)^n
\left(c\e^{r_0}\|\Psi\|_{r_0}|\Lambda|^{-1/d}n^d+b_\Lambda\right)
\rightarrow0
$$
along the centered cubes, provided that the series converges. Since $\bar t<T_\theta/2$ implies
$$
\tilde C
\leq
\frac2{1-\theta/\Theta_\Psi},
$$
it is enough to require
$$
\bar t<T_2\coloneq
\min\left\{
\frac{T_\theta}{2},
\frac{\Theta_\Phi}{4}\e^{-(\rho+\eta)\kappa_R}
\left(1-\frac{\theta}{\Theta_\Psi}\right)^2
\right\}.
$$
The estimates also establish locally uniform convergence on
$|\Im z|<\theta$ of both series obtained by termwise analytic
continuation. For real $z$, their sums agree with the corresponding
real-time evolutions of $W_\Lambda(t)$, and hence provide the
required analytic extensions. Thus, the assertions hold with
$\bar T_\theta\coloneq\min\{T_1,T_2\}$. \qed

Let $\omega\in\cSwg(\Psi)$. By the choice of $\rho$, 
$\Theta_\Psi>\frac12$. Choose $\frac12<\theta<\Theta_\Psi$. For
$|t|<\bar T_\theta$, Theorem~\ref{thm:alphaW_final} verifies the assumptions
of Theorem~\ref{thm:WG}, and hence $\omega_t\in\cSwg(H(t))$. This proves
Part~\textup{(i)} of Theorem~\ref{thm:wgPreserv}.

\subsubsection{Proof of Part~(ii)}
\label{sec:WG_1dim}

Throughout this section we assume that  $d=1$ and that $\Phi,\Psi\in\cBf$. 
Let $R$ and $R'$ be the ranges of $\Psi$ and $\Phi$, respectively.

To apply Araki's estimates uniformly to restricted interactions,
we first regroup the interaction terms into operators supported
on intervals of fixed length. Let $\Theta\in\cBf$ have range $L$,
and set
$$
\II_L\coloneq\llbracket0,L\rrbracket,\qquad
\II_L^x\coloneq\llbracket x,x+L\rrbracket,
$$
for $x\in\ZZ$. For any non-empty $X\in\cF$ with $\diam(X)\leq L$, let
$$
N_L(X)\coloneq
\left|
\left\{
x\in\ZZ\mid X\subset\II_L^x
\right\}
\right|
=
L-\diam(X)+1.
$$
Given $K\subset\ZZ$ and $x\in\ZZ$, define
$$
\widehat{\Theta}_{K,x}
\coloneq
\sum_{\substack{\emptyset\neq X\subset K\cap\II_L^x}}
\frac{\Theta(X)}{N_L(X)}
\in\fA_{K\cap\II_L^x}.
$$
Denote by $D_{\widehat{\Theta}_K}(x)$ the $\ast$-derivation
$$
D_{\widehat{\Theta}_K}(x)A
\coloneq
\i[\widehat{\Theta}_{K,x},A],
$$
and, for $\by=(y_1,\ldots,y_m)\in\ZZ^m$, set
$$
D_{\widehat{\Theta}_K}(\by)
\coloneq
D_{\widehat{\Theta}_K}(y_m)\cdots
D_{\widehat{\Theta}_K}(y_1).
$$

\bel\label{lem:delta_Araki}
For any $K\subset\ZZ$ and $x\in\ZZ$, one has
$$
\|\widehat{\Theta}_{K,x}\|
\leq
(L+1)\|\Theta\|_\mathrm{s}.
$$
Moreover, for every $A\in\fAloc$,
\beq
\delta_{\Theta|_K}(A)
=
\sum_{x\in\ZZ}D_{\widehat{\Theta}_K}(x)A.
\label{eq:restricted_block_decomposition}
\eeq
\eel

\proof
The first assertion follows from
\[
\|\widehat{\Theta}_{K,x}\|
\leq
\sum_{\emptyset\neq X\subset\II_L^x}\|\Theta(X)\|
\leq
\sum_{y\in\II_L^x}\sum_{X\ni y}\|\Theta(X)\|
=
(L+1)\|\Theta\|_\mathrm{s}.
\]
Since every non-empty $X$ with $\diam(X)\leq L$ is contained in exactly
$N_L(X)$ intervals of the form $\II_L^x$, we further have
\[
\sum_{x\in\ZZ}D_{\widehat{\Theta}_K}(x)A
=
\sum_{x\in\ZZ}
\sum_{\substack{\emptyset\neq X\subset K\cap\II_L^x}}
\frac{\i[\Theta(X),A]}{N_L(X)}
=
\sum_{\substack{\emptyset\neq X\subset K\\X\cap\supp A\neq\emptyset}}
\i[\Theta(X),A]
=
\delta_{\Theta|_K}(A),
\]
which proves~\eqref{eq:restricted_block_decomposition}.
\qed

Let $J\subset\ZZ$ be a finite interval. For $m\geq1$, we denote by
$\cC'_{L,m}(J)$ the set of all $\by=(y_1,\ldots,y_m)\in\ZZ^m$ such that
$$
\II_L^{y_j}\cap
\left(
J\cup\bigcup_{k=1}^{j-1}\II_L^{y_k}
\right)
\neq\emptyset
$$
for every $j\in\llbracket1,m\rrbracket$. If
$J=\llbracket0,n\rrbracket$, we write $\cC'_{L,m}(n)$ instead of
$\cC'_{L,m}(J)$, and  set
$$
\cC'_{L,0}(J)=\cC'_{L,0}(n)=\{\emptyset\},
\qquad
D_{\widehat{\Theta}_K}(\emptyset)=\Id.
$$

\bel\label{lem:HHhat}
\benr
\item If $n\geq L-1$, then, for every $\theta\geq0$,
$$
F_{L,n}(\theta)
\coloneq
\sum_{m\geq0}\frac{\theta^m}{m!}|\cC'_{L,m}(n)|
=
\exp\left(
(n-L+1)\theta
+
2\sum_{q=1}^L\frac{\e^{q\theta}-1}{q}
\right).
$$

\item Let $K\subset\ZZ$, $A\in\fAloc$, and let $J$ be a finite interval
containing $K\cap\supp A$. Then, for every $m\geq0$,
\beq
\delta_{\Theta|_K}^m(A)
=
\sum_{\by\in\cC'_{L,m}(J)}
D_{\widehat{\Theta}_K}(\by)A.
\label{eq:restricted_block_iterates}
\eeq
\eenr
\eel

\proof
\textbf{(i)} This follows from~\cite[Lemma~4.1 and the proof of Theorem~4.2]{Araki1969}.

\medskip\noindent\textbf{(ii)}
Iterating~\eqref{eq:restricted_block_decomposition} gives
$$
\delta_{\Theta|_K}^m(A)
=
\sum_{\by\in\ZZ^m}D_{\widehat{\Theta}_K}(\by)A.
$$
If a term in this sum does not vanish, then the support of
$\widehat{\Theta}_{K,y_1}$ intersects $K\cap\supp A\subset J$ and, successively,
the support of $\widehat{\Theta}_{K,y_j}$ intersects
$$
\supp A\cup
\bigcup_{k=1}^{j-1}\supp\widehat{\Theta}_{K,y_k}.
$$
Since $\supp\widehat{\Theta}_{K,y_k}\subset\II_L^{y_k}$, 
it follows that $\by\in\cC'_{L,m}(J)$, which proves
\eqref{eq:restricted_block_iterates}.
\qed

\bep\label{prop:arakis_bound_dim1}
Let $\Theta\in\cBf$ have range $L$, and set
$$
C_\Theta\coloneq2(L+1)\|\Theta\|_\mathrm{s}.
$$
Let $K\subset\ZZ$, $A\in\fAloc$, and let $n\geq L-1$ be such that
$K\cap\supp A$ is contained in an interval of length $n+1$. Then the map
$$
\RR\ni s\mapsto\tau_{\Theta|_K}^s(A)
$$
has an entire analytic extension and, for every $z\in\CC$,
\beq
\|\tau_{\Theta|_K}^z(A)\|
\leq
F_{L,n}(C_\Theta|\Im z|)\|A\|.
\label{bound-knee}
\eeq
\eep

\proof Let $J$ be an interval of length $n+1$ containing $K\cap\supp A$.
By Lemma~\ref{lem:HHhat},
\[
\|\delta_{\Theta|_K}^m(A)\|
\leq
\sum_{\by\in\cC'_{L,m}(J)}
\|D_{\widehat{\Theta}_K}(\by)A\|
\leq
2^m\|A\|
\sum_{\by\in\cC'_{L,m}(J)}
\prod_{j=1}^m\|\widehat{\Theta}_{K,y_j}\|
\leq
C_\Theta^m|\cC'_{L,m}(n)|\|A\|,
\]
where $|\cC'_{L,m}(J)|=|\cC'_{L,m}(n)|$ by translation. Hence, the Taylor series
$$
\sum_{m\geq0}\frac{z^m}{m!}\delta_{\Theta|_K}^m(A)
$$
converges in norm for every $z\in\CC$,  and
defines the entire analytic extension of $\tau_{\Theta|_K}^s(A)$. For
$z\in\i\RR$, Lemma~\ref{lem:HHhat}(i) gives
$$
\|\tau_{\Theta|_K}^z(A)\|
\leq
F_{L,n}(C_\Theta|z|)\|A\|.
$$
The bound~\eqref{bound-knee} follows from the group property and the isometry of the
real-time dynamics.
\qed

Set
$$
C_\Psi\coloneq2(R+1)\|\Psi\|_\mathrm{s},
\qquad
C_\Phi\coloneq2(R'+1)\|\Phi\|_\mathrm{s}.
$$
For $t\in\RR$, $u\geq0$, and $n\geq\max\{R-1,R'\}$, define
$$
G(t,u,n)
\coloneq
F_{R,n}(C_\Psi u)
F_{R',n}
\left(
C_\Phi|t|F_{R,n}(C_\Psi u)
\right).
$$

\bep\label{prop:arakis_bound}
Let $t\in\RR$ and $K\subset\ZZ$. Let $A\in\fAloc$, and let
$n\geq\max\{R-1,R'\}$ be such that $K\cap\supp A$ is contained in an
interval of length $n+1$. Then the map
$$
\RR\ni s\mapsto\alpha_K^s(A)
$$
has an entire analytic extension and, for every $z\in\CC$,
$$
\|\alpha_K^z(A)\|
\leq
G(t,|\Im z|,n)\|A\|.
$$
\eep
\proof Choose an interval $J$ of length $n+1$ containing $K\cap\supp A$.
By Lemma~\ref{lem:HHhat},
$$
\tau_{\Phi|_K}^t(A)
=
\sum_{m\geq0}\frac{t^m}{m!}
\sum_{\by\in\cC'_{R',m}(J)}
D_{\widehat{\Phi}_K}(\by)A.
$$
For real $s$, we apply $\tau_{\Psi|_K}^s$ termwise and then
$\tau_{\Phi|_K}^{-t}$. The estimates below show that the resulting
series, with $s$ replaced by $z$, converges locally uniformly on $\CC$
and defines the entire extension $\alpha_K^z(A)$.

Set
$$
f(z)\coloneq F_{R,n}(C_\Psi|\Im z|).
$$
Proposition~\ref{prop:arakis_bound_dim1} gives
$$
\|\tau_{\Psi|_K}^z(A)\|
\leq
f(z)\|A\|,
\qquad
\|\tau_{\Psi|_K}^z(\widehat{\Phi}_{K,y})\|
\leq
\frac{C_\Phi}{2}f(z).
$$
These bounds yield
\[
\left\|
\tau_{\Psi|_K}^z
\left(
D_{\widehat{\Phi}_K}(\by)A
\right)
\right\|\leq
2^m
\|\tau_{\Psi|_K}^z(A)\|
\prod_{j=1}^m
\|\tau_{\Psi|_K}^z(\widehat{\Phi}_{K,y_j})\|\leq
C_\Phi^m f(z)^{m+1}\|A\|.
\]
Since $\tau_{\Phi|_K}^{-t}$ is isometric, it follows that
\begin{align*}
\|\alpha_K^z(A)\|
&\leq
f(z)
\sum_{m\geq0}
\frac{\left(C_\Phi|t|f(z)\right)^m}{m!}
|\cC'_{R',m}(n)|\,\|A\|\\
&=
F_{R,n}(C_\Psi|\Im z|)
F_{R',n}
\left(
C_\Phi|t|F_{R,n}(C_\Psi|\Im z|)
\right)\|A\|,
\end{align*}
as claimed.
\qed

\ber
For fixed $t\in\RR$ and $u\geq0$, the function $n\mapsto G(t,u,n)$ is non-decreasing.
\eer

Finally, we establish the uniform bounds on the evolved surface energies.

\bet\label{thm:d1_extension}
Let $t\in\RR$. For any interval
$\Lambda=\llbracket a,b\rrbracket\subset\ZZ$, the maps
$$
\RR\ni s\mapsto\alpha^s(W_\Lambda(t)),
\qquad
\RR\ni s\mapsto
\alpha^s_{-W_\Lambda(t)}(W_\Lambda(t))
$$
have entire analytic extensions. Moreover, for every $a_0>0$ there exists a
constant $C_{t,a_0}$, independent of $\Lambda$, such that
$$
\sup_{|\Im z|<a_0}
\left(
\|\alpha^z(W_\Lambda(t))\|
+
\|\alpha^z_{-W_\Lambda(t)}(W_\Lambda(t))\|
\right)
\leq
C_{t,a_0}.
$$
\eet

\proof
Set $n_0\coloneq\max\{R,R'\}$. 
For $X\in\cF$, $K\subset\ZZ$, and $m\geq0$, define
$$
Q_K^{(m)}(X)
\coloneq
\delta_{\Phi|_K}^m(\Psi(X))
=
\sum_{\bY\in\cC_m^K(X)}D_\Phi(\bY)\Psi(X).
$$
If $\Psi(X)\neq0$, choose an interval $J_X$ of length $n_0+1$ containing
$X$. Lemma~\ref{lem:HHhat} gives
\beq
Q_K^{(m)}(X)
=
\sum_{\by\in\cC'_{R',m}(J_X)}
D_{\widehat{\Phi}_K}(\by)\Psi(X).
\label{eq:Q_block_expansion}
\eeq
For an interval $\Lambda\subset\ZZ$, set
$$
B_\Lambda^{(m)}(X)
\coloneq
\sum_{\bY\in\cC_m(X,\partial\Lambda)}
D_\Phi(\bY)\Psi(X).
$$
By~\eqref{eq:need_chains4},
\beq
B_\Lambda^{(m)}(X)
=
\begin{cases}
Q_\ZZ^{(m)}(X)-Q_\Lambda^{(m)}(X)
&\text{if }X\subset\Lambda;\\[4pt]
Q_\ZZ^{(m)}(X)-Q_{\Lambda^c}^{(m)}(X)
&\text{if }X\subset\Lambda^c;\\[4pt]
Q_\ZZ^{(m)}(X)
&\text{if }X\cap\Lambda\neq\emptyset\neq X\cap\Lambda^c.
\end{cases}
\label{eq:boundary_Q_decomposition}
\eeq
For $|t|<\Theta_\Phi$, Lemma~\ref{lem:WG_formula} and
\eqref{eq:boundary_Q_decomposition} give
\beq
W_\Lambda(t)
=
\sum_{m\geq0}\frac{(-t)^m}{m!}
\sum_{X\in\cF}B_\Lambda^{(m)}(X).
\label{eq:formula_Wt_1d}
\eeq
We proceed to  show that the series on the right-hand side converges
locally uniformly in $t\in\CC$ and hence defines an entire $\fA$-valued
function.

If $B_\Lambda^{(m)}(X)\neq0$ and $\Psi(X)\neq0$, then
$$
X\cap\cT_m^\Lambda\neq\emptyset,
\qquad
\cT_m^\Lambda
\coloneq
\left\{
x\in\ZZ\mid
\dist(\{x\},\partial\Lambda)\leq R+mR'
\right\}.
$$
There exists a constant $c>0$, independent of $\Lambda$ and $m$, such that
\beq
|\cT_m^\Lambda|
\leq
c(1+m).
\label{eq:one_dim_tube_bound}
\eeq
Moreover, Lemma~\ref{lem:delta_Araki} and
\eqref{eq:Q_block_expansion} give
$$
\|Q_K^{(m)}(X)\|
\leq
C_\Phi^m
|\cC'_{R',m}(n_0)|
\|\Psi(X)\|,
$$
and hence
$$
\|B_\Lambda^{(m)}(X)\|
\leq
2C_\Phi^m
|\cC'_{R',m}(n_0)|
\|\Psi(X)\|.
$$
It follows that
\beq
\sum_{m\geq0}\frac{|t|^m}{m!}
\sum_{X\in\cF}\|B_\Lambda^{(m)}(X)\|
\leq
2c\|\Psi\|_\mathrm{s}
\sum_{m\geq0}
(1+m)
\frac{(C_\Phi|t|)^m}{m!}
|\cC'_{R',m}(n_0)|
<\infty,
\label{mar-phy-bb}
\eeq
where the last assertion follows from Lemma~\ref{lem:HHhat}(i). The same estimates apply to the Taylor series of
$W_{\Lambda,\widetilde\Lambda}(t)$, uniformly in
$\widetilde\Lambda\supset\Lambda$. Since the interactions have finite
range, each coefficient agrees with the corresponding coefficient
in~\eqref{eq:formula_Wt_1d} once $\widetilde\Lambda$ is sufficiently
large. The  bound~\eqref{mar-phy-bb} gives locally uniform convergence on
$\CC$ to the right-hand side of~\eqref{eq:formula_Wt_1d}.
For real $t$, Proposition~\ref{prop:formula_W} identifies this limit
with $W_\Lambda(t)$. Thus~\eqref{eq:formula_Wt_1d} holds for every
$t\in\RR$ and defines an entire $\fA$-valued extension of
$t\mapsto W_\Lambda(t)$.

Fix $a_0>0$ and set $g\coloneq G(t,a_0,n_0)$. 
For $|\Im z|<a_0$, Proposition~\ref{prop:arakis_bound} and
\eqref{eq:Q_block_expansion} give
\beq
\|\alpha^z(Q_K^{(m)}(X))\|
\leq
C_\Phi^m
|\cC'_{R',m}(n_0)|
g^{m+1}
\|\Psi(X)\|,
\label{eq:Q_alpha_bound}
\eeq
uniformly in $K\subset\ZZ$. Combining
\eqref{eq:formula_Wt_1d},
\eqref{eq:boundary_Q_decomposition},
\eqref{eq:one_dim_tube_bound}, and
\eqref{eq:Q_alpha_bound}, we obtain
\begin{align*}
\|\alpha^z(W_\Lambda(t))\|
&\leq
2cg\|\Psi\|_\mathrm{s}
\sum_{m\geq0}
(1+m)
\frac{(C_\Phi g|t|)^m}{m!}
|\cC'_{R',m}(n_0)|.
\end{align*}
The bound is finite and uniform in $\Lambda$ and $|\Im z|<a_0$.
Since $a_0>0$ is arbitrary, the series
$$
\alpha^z(W_\Lambda(t))
=
\sum_{m\geq0}\frac{(-t)^m}{m!}
\sum_{X\in\cF}\alpha^z(B_\Lambda^{(m)}(X))
$$
converges locally uniformly on $\CC$ and provides the required entire
extension.

For a non-empty finite set $Y\subset\ZZ$, set
$$
n(Y)\coloneq\max\{\diam(Y),R-1,R'\},
\qquad
\chi_{a_0}(Y)\coloneq G(t,a_0,n(Y)),\qquad \chi_{a_0}(\emptyset)=1.
$$
By the preceding remark,
$\chi_{a_0}$ is non-decreasing. Proposition~\ref{prop:arakis_bound}
therefore verifies the assumptions of
Lemma~\ref{lem:restricted_double_ineq} on the strip
$|\Im z|<a_0$. In particular,
$$
\|\alpha^z_{-W_\Lambda(t)}(\Psi(X))\|
\leq
g^2\|\Psi(X)\|,
\qquad
\|\alpha^z_{-W_\Lambda(t)}(\widehat{\Phi}_{K,y})\|
\leq
\frac{C_\Phi}{2}g^2.
$$
Using again
\eqref{eq:Q_block_expansion}, we obtain
$$
\|\alpha^z_{-W_\Lambda(t)}(Q_K^{(m)}(X))\|
\leq
C_\Phi^m
|\cC'_{R',m}(n_0)|
g^{2m+2}
\|\Psi(X)\|,
$$
uniformly in $K$ and $\Lambda$. Consequently,
\begin{align*}
\|\alpha^z_{-W_\Lambda(t)}(W_\Lambda(t))\|
&\leq
2cg^2\|\Psi\|_\mathrm{s}
\sum_{m\geq0}
(1+m)
\frac{(C_\Phi g^2|t|)^m}{m!}
|\cC'_{R',m}(n_0)|.
\end{align*}
The estimate shows that applying $\alpha_{-W_\Lambda(t)}^z$
termwise to~\eqref{eq:formula_Wt_1d} gives an operator series
converging uniformly for $|\Im z|<a_0$, with its sum bounded
independently of $\Lambda$. For real $z=s$, this sum equals
$\alpha_{-W_\Lambda(t)}^s(W_\Lambda(t))$. Since $a_0>0$ is
arbitrary, the series provides  the required entire extension.
\qed

Let $\omega\in\cSwg(\Psi)$. Taking
$a_0>1/2$ in Theorem~\ref{thm:d1_extension}, we obtain
$$
\Llim\frac1{|\Lambda|}
\left(
\|\alpha^{\i/2}(W_\Lambda(t))\|
+
\|\alpha^{\i/2}_{-W_\Lambda(t)}(W_\Lambda(t))\|
\right)
=0.
$$
Theorem~\ref{thm:WG} then  gives $\omega_t\in\cSwg(H(t))$ for all
$t\in\RR$. This proves Part~\textup{(ii)} of
Theorem~\ref{thm:wgPreserv}.

\printbibliography

\end{document}